\documentclass[manuscript]{aastex}
\usepackage{bm}
\usepackage{url}
\usepackage{amsmath}

\newcommand{\myemail}{t.ueda@geo.titech.ac.jp}

\shorttitle{}
\shortauthors{Ueda et al.}

\begin{document}

\title{
Size Dependence of Dust Distribution around the Earth Orbit
}


\author{Takahiro Ueda \altaffilmark{1},}
\email{\myemail}
\author{Hiroshi Kobayashi \altaffilmark{2},} 
\author{Taku Takeuchi \altaffilmark{1,3},}
\author{Daisuke Ishihara \altaffilmark{2},} 
\author{Toru Kondo\altaffilmark{2},}
\and
\author{Hidehiro Kaneda \altaffilmark{2}}
\altaffiltext{1}{Department of Earth and Planetary Sciences, Tokyo Institute of Technology, Meguro, Tokyo, 152-8551, Japan}
\altaffiltext{2}{Graduate School of Science, Nagoya University, Nagoya, Aichi, 464-8602, Japan}
\altaffiltext{3}{Present affiliation: Sanoh Industrial Co., Ltd.}

\begin{abstract}
In the Solar System, interplanetary dust particles (IDPs) originating mainly from asteroid collisions and cometary activities drift to the Earth orbit due to the Poynting-Robertson drag.
We analyzed the thermal emission from IDPs that was observed by the first Japanese infrared astronomical satellite, AKARI. 
The observed surface brightness in the trailing direction of the Earth orbit is 3.7\% greater than that in the leading direction in the $9\,\micron$ band and 3.0\% in the $18\,\micron$ band.
In order to reveal dust properties causing the leading-trailing surface brightness asymmetry, we numerically integrated orbits of the Sun, the Earth, and a dust particle as a restricted three-body problem including radiation from the Sun. 
The initial orbits of particles are determined according to the orbits
 of main-belt asteroids or Jupiter-family comets.  The orbital trapping
 in mean motion resonances results in a significant leading-trailing asymmetry
 so that intermediate sized dust ($\sim 10$--$100\,\micron$) produces a greater asymmetry than the zodiacal light has.
The leading-trailing surface brightness difference integrated over the size distribution of the asteroidal dust is obtained to be the values of 27.7\% and 25.3\% in the $9\,\micron$ and $18\,\micron$ bands, respectively. 
In contrast, the brightness difference for cometary dust is calculated as the values of 3.6\% and 3.1\% in the $9\,\micron$ and 18\,$\micron$ bands, respectively, if the maximum dust radius is set to be $s_{\rm max}=3000\,\micron$.
Taking into account these values and their errors, we conclude that the contribution of asteroidal dust to the zodiacal infrared emission is less than
 $\sim 10\%$, while cometary dust of the order of 1\,mm mainly accounts for the zodiacal light in infrared. 

\end{abstract}

\keywords{interplanetary medium --- zodiacal dust --- comets: general --- minor planets, asteroids: general}

\section{Introduction}
\label{sec:intro}
In our solar system, there are many dust particles originating from
asteroid collisions, cometary activities, interstellar space and so on.  These
dust particles gradually drift to around the Earth because they lose
their angular momentum due to the absorption and re-radiation of the
sunlight, which is called Poynting-Robertson effect (hereafter P-R
effect; e.g., \citealt{Burns1979}).  The thermal emission and scattering
light from the drifting particles are called the ``zodiacal light''. 
The spatial distribution of the zodiacal light depends on the orbital evolution
of drifting dust particles, which is characterized by the parameter
$\beta$ that represents the ratio of solar radiation pressure to solar gravity.
The parameter $\beta$ is defined as a function of dust properties such as dust radius $s$ and material
density $\rho_{s}$.  Therefore, investigating $\beta$ from the spatial
distribution of the zodiacal light gives constraints on the properties
of the zodiacal dust particles, which may reveal the properties of
parent bodies and the origin of them.

\begin{figure}[h]
\begin{center}
\includegraphics[scale=0.42]{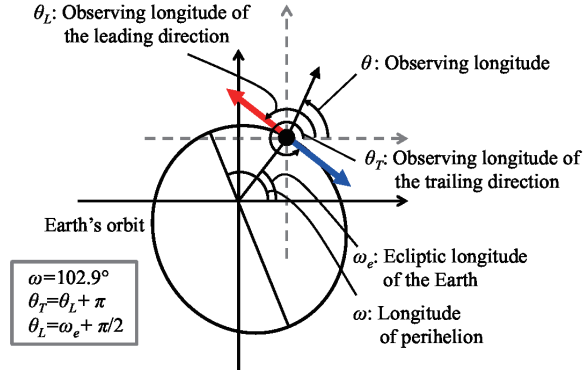}
\caption{
The definitions of the various sorts of the angles and the relationship between each angles.
}
\label{fig:fig}
\end{center}
\end{figure}

The interplanetary dust cloud is mainly composed of three components, smooth cloud that is the main contributor to the zodiacal light, asteroidal dust bands, and circumsolar ring caused by dust particles trapped in mean motion resonances (MMRs) with the Earth.
The Infrared Astronomical Satellite (IRAS) performed an all-sky survey
of 4 infrared bands centered at wavelengths 12, 25, 60, and 100$\,\micron$. 
The observations of the zodiacal light with IRAS showed that brightnesses in the trailing direction of the Earth are 3--4\% greater than those in the leading direction in the 12, 25, and $60\,\micron$ bands (\citealt{Dermott1994}, 
the definitions of the leading and trailing directions are shown in Figure \ref{fig:fig}).
\citet{Dermott1994} also carried out orbital calculations for asteroidal dust particles with $s=6\,\micron$ and $\rho_{s}=2.7\,{\rm g/cm^{3}}$ 
(corresponding to $\beta=0.037$) and empirically derived the relative probability of capture into and the average trapping time in
$p:(p+1)$ MMRs outside the orbit of the Earth. 
\citet{Dermott1994} made a model ``image'' of a circumsolar ring in the zodiacal
dust cloud using these empirical data and showed that the leading-trailing brightness asymmetry
observed by IRAS is caused by the circumsolar ring composed of asteroidal dust (see also \citealt{Jackson1989}).  
However, \citet{Dermott1994} carried out these simulations only for asteroidal particles with $\beta=0.037$.  
One of the our purposes is to investigate the dependence of the leading-trailing surface brightness asymmetry on $\beta$ and the origin of dust particles.

The COBE Diffuse Infrared Background Experiment (DIRBE) also
investigated the zodiacal light flux in each of the 10 wavelength bands
ranging from $1.25$ to $240\,\micron$.  \citet{Reach1995} confirmed the existence of the ring structure predicted
by \citet{Dermott1994} using the observations with COBE.  From COBE
observations, \citet{Kelsall1998} developed a density distribution model of
the zodiacal cloud by parametric fitting with more than $30$ parameters
(in addition, there are also $\sim$40 source function parameters).  An
improved model for the zodiacal cloud infrared emission was constructed
by \citet{Rowan2013} using not only the observations of COBE
but also those of IRAS.

The origin of zodiacal dust is one of the most difficult subjects and
still under discussion.  \citet{Liou1995} studied the orbital
evolution of dust particles with $\beta=0.05037$ originating from
Encke-type comets by numerical integration. 
Cometary dust particles are found to have a broad vertical distribution due to their large inclinations.
\citet{Liou1995} concluded that the latitude profiles of the zodiacal
emission observed with IRAS can be accounted for by a combination of
$\sim$ 1/4--1/3 asteroidal particles and $\sim$ 3/4--2/3 Encke-type cometary particles. 
Based on modeling similar to that of \citet{Liou1995}, 
\citet{Nesvorny2010} found that 85--95\% of the observed mid-infrared emission
is produced by particles from Jupiter-Family comets and $<10\%$ from asteroids.
The optical properties such as albedo and spectral gradient of the zodiacal light are explained if more than 90\% of the zodiacal dust originates
from comets or D-type asteroids \citep{Yang2015}.
The zodiacal dust from other origins is considered to be minor.  
The dust particles produced in the Kuiper belt are estimated to be as low as 1--2\% of IDPs in number \citep{Moro-martin2003}.  
Although these studies suggest that cometary dust is a main component of the zodiacal cloud, the population coming from the asteroids or comets is still under debate.

The size distribution of IDPs is also a difficult subject because sensitive size ranges are different for different
measurement methods and we have to comprehensively consider results of various methods (see \citealt{Sykes2004}).  
One of the most well-known size distribution model was proposed by \citet{Grun1985}, who interpolated the distribution of large meteoroids ($m \gtrsim 10^{-6}\,{\rm g}$) derived from the lunar crater size distribution and that of small meteoroids ($m \lesssim 10^{-9}\,{\rm g}$)
derived from the meteoroid flux measurements with the HEOS-2 and Pegasus satellite and a theoretical model of small particles.  
According to this model, the cross-sectional distribution of IDPs has a
peak at $m \sim 3 \times 10^{-7}\,{\rm g}$ that corresponds to $s \sim 33\,\micron$ with $\rho_{s}=2\,{\rm g/cm^{3}}$ (see Figure 4 of \citealt{Grun1985}). 
\citet{Love1993} derived the mass flux distribution using a direct experiment measuring hypervelocity impact
craters on Long Duration Exposure Facility (LDEF) satellite. 
The mass flux distribution obtained by \citet{Love1993} peaks near $m \sim1.5 \times 10^{-5}\,{\rm g}$ corresponding to
a peak at $s \sim 121\,\micron$ with $\rho_{s}=2\,{\rm g/cm^{3}}$ in the cross-sectional distribution.
This peak size is 3.7 times larger than that of \citet{Grun1985} (see also \citealt{Farley1997}).
However, the size distribution of cometary dust, which is thought to be a main component
of zodiacal dust, is different from these expected from near-Earth
direct measurements.  NASA's Stardust mission carried out the direct
measurement of cometary dust in the coma of comet 81P/Wild 2 (e.g., \citealt{Horz2006} \citealt{Price2010}). 
The dust particles originating from comet 81P/Wild 2 have a
differential number density distribution of $n_{\rm d}(s) \propto s^{-2.72}$ \citep{Horz2006}. 
The direct measurements within the inner coma of comet 1P/Halley also
suggest that larger particles dominate the thermal emission from the
cometary dust particles (e.g., \citealt{McDonnell1987}, \citealt{Kolokolova2007}). 
The most recent in-situ observation, Rosetta mission, showed that the dust particles originating from comet 67P/Churyumov-Gerasimenko
have the distribution of $n_{\rm d}(s) \propto s^{-3}$ for particle smaller than
$\sim 1\,{\rm mm}$ \citep{Moreno2016}.
According to the size distribution of cometary dust, the total mass and cross-sectional area of dust particles are dominated by millimeter sized particles, which seems different from the size distribution of IDPs expected from near-Earth direct measurements shown by \citet{Grun1985} and \citet{Love1993}.
In order to clarify the size distribution, it is worthwhile determining $\beta$ from the leading-trailing brightness difference.


The first Japanese infrared astronomical satellite AKARI have performed an all-sky survey of 6 infrared bands centered at wavelengths 9, 18, 65, 90, 140 and 160$\,\micron$ from May 2006 to August 2007 with higher spatial resolutions than COBE and IRAS (e.g., \citealt{Pyo2010}, \citealt{Doi2015}, \citealt{Kondo2016}).
AKARI revolved around the Earth in a sun-synchronous polar orbit and scanned the sky along the circle of the solar elongation at approximately $90^{\circ}$.
The asymmetry in the leading and trailing brightnesses obtained from the AKARI observations in the $9\,\micron$ and $18\,\micron$ bands mainly comes from the thermal emission of dust grains around the Earth \citep{Kondo2016}.
In order to reveal dust properties causing the leading-trailing brightness asymmetry, we conduct a number of orbital calculations of a dust particle originating from main-belt asteroids and Jupiter-family comets with a broad range of $\beta$.

In Section \ref{sec:obs}, we present the analysis of the AKARI observations in the $9\,\micron$ and $18\,\micron$ bands.
We introduce our orbital calculation method in Section \ref{sec:model} and show the spatial distribution of grains with each $\beta$ in Section \ref{sec:orbital}.
In Section \ref{sec:discuss}, we discuss the leading-trailing brightness difference taking into account the size distributions of asteroidal and cometary dust and the perturbation from planets other than the Earth.
We summarize our findings in Section \ref{sec:summary}.

\section{Analysis of the AKARI Observations}
\label{sec:obs}
\begin{figure*}[h]
\begin{center}
\includegraphics[scale=0.55]{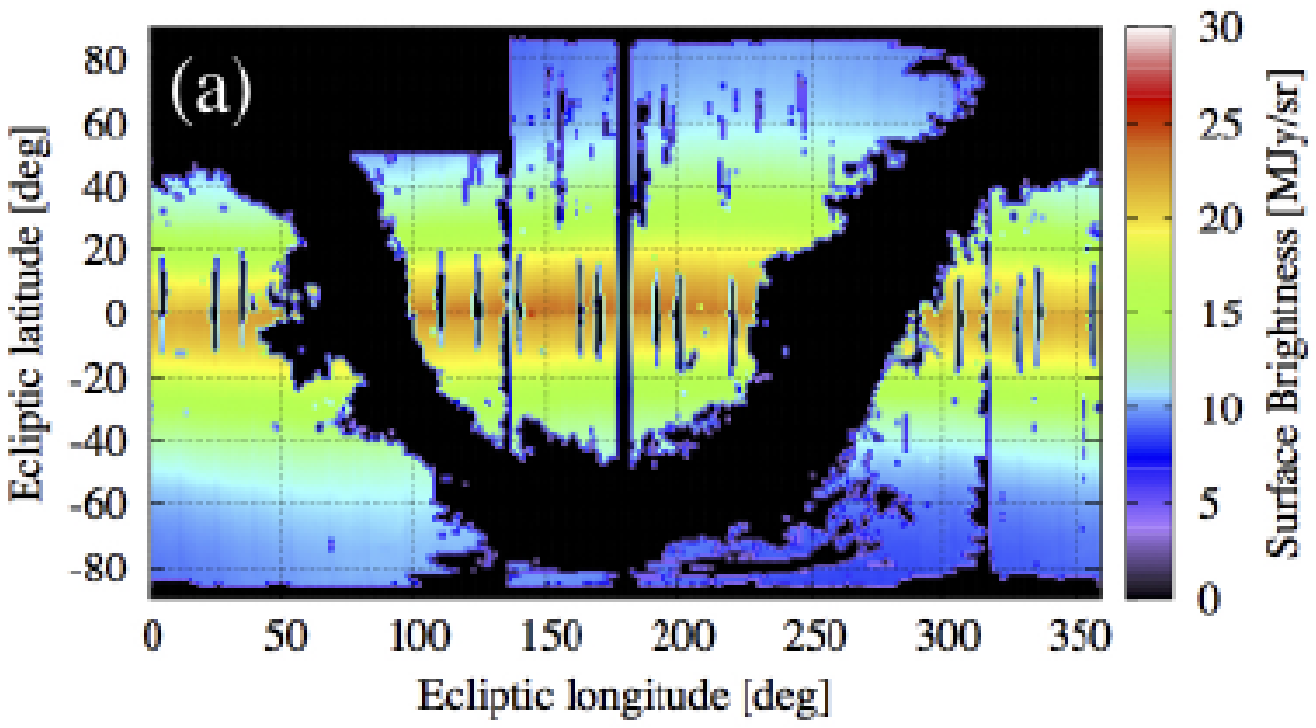}
\label{fig:lS}
\includegraphics[scale=0.55]{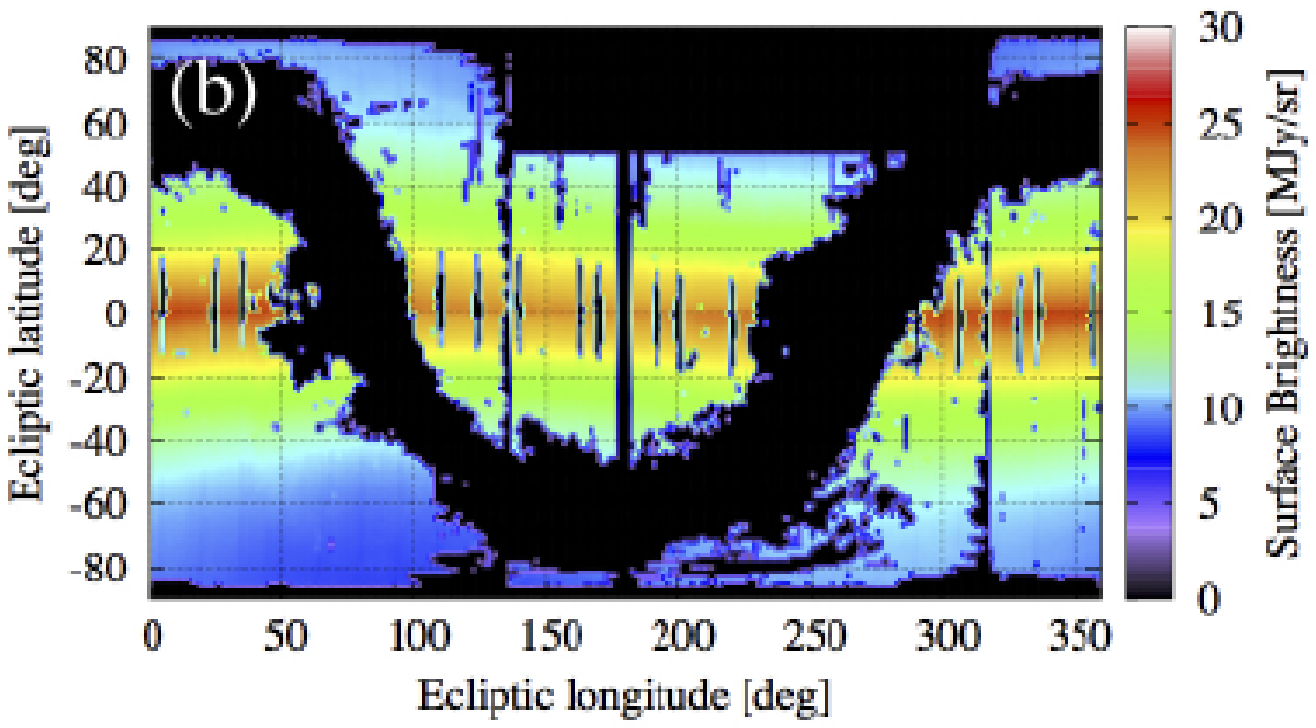}
\label{fig:tS}
\\
\includegraphics[scale=0.55]{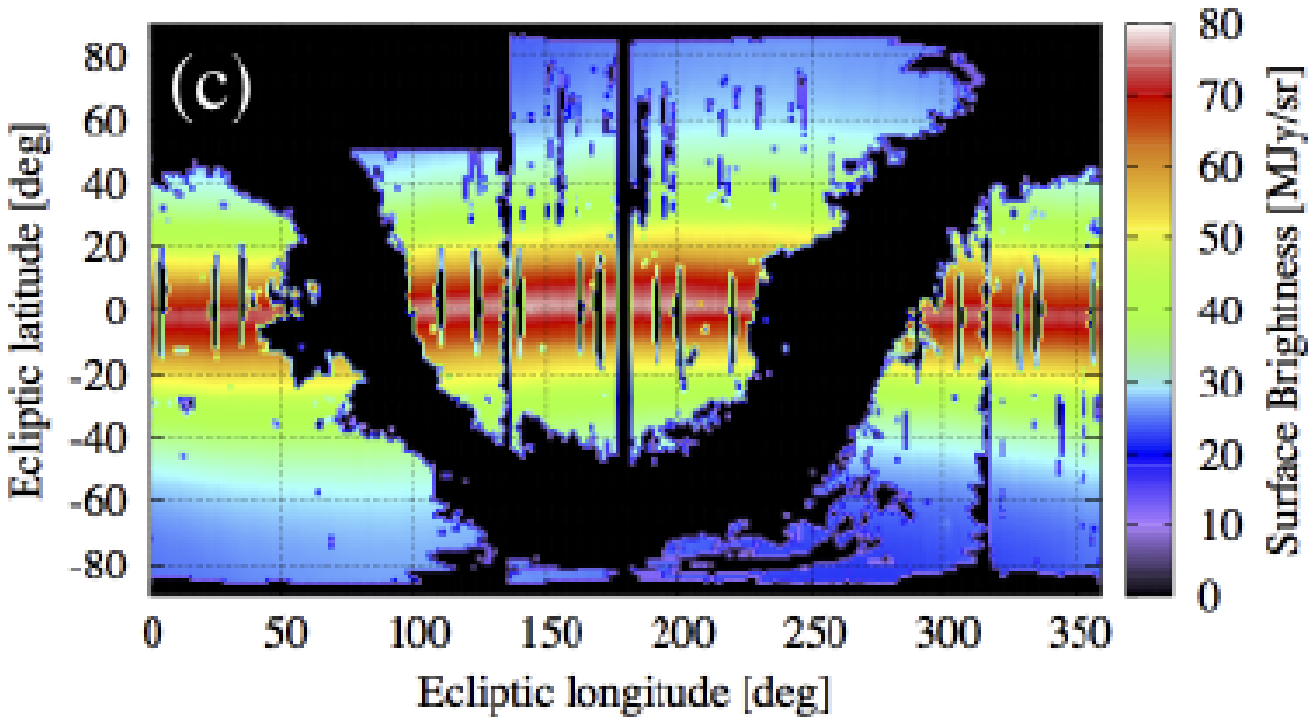}
\label{fig:lL}
\includegraphics[scale=0.55]{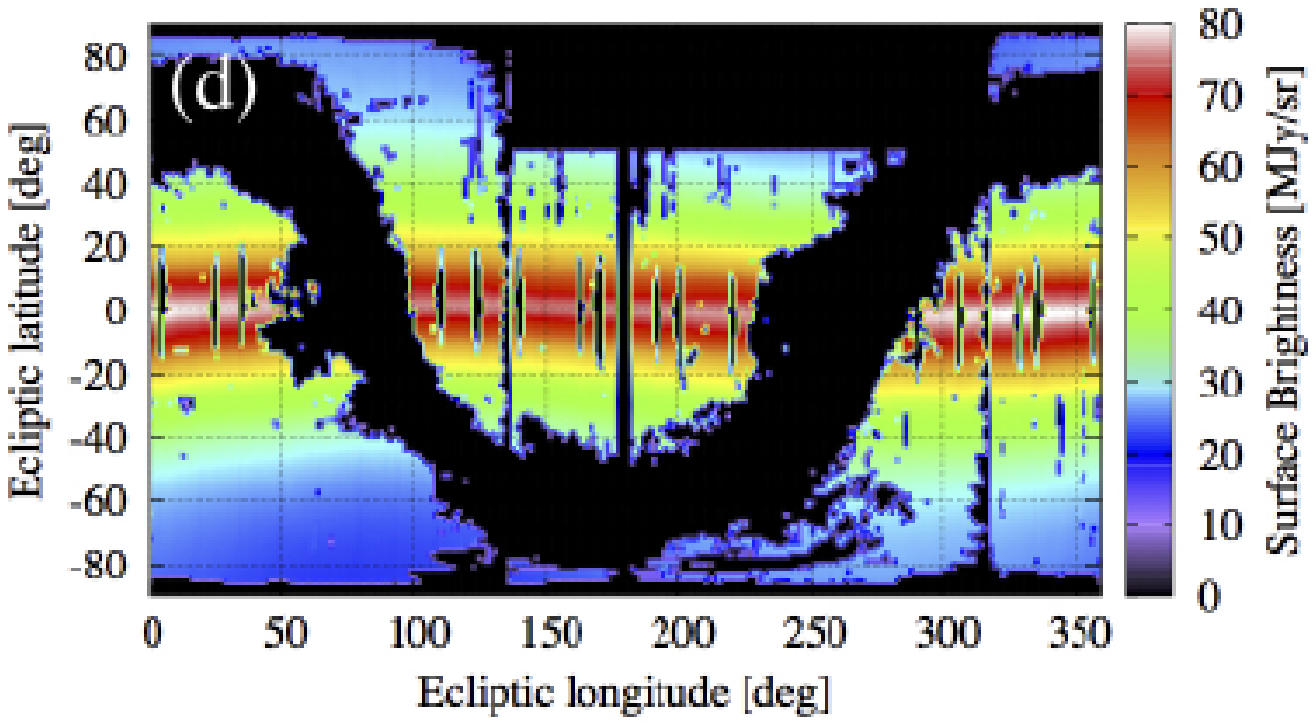}
\label{fig:tL}
\caption{
AKARI all sky survey maps in the $9\,\micron$ band (top panels) and $18\,\micron$ band (bottom panels).
Left-hand panels show the surface brightnesses observed in leading direction of the Earth and right-hand panels show those in trailing direction.
}
\label{fig:akari_map}
\end{center}
\end{figure*}
\begin{figure}[h]
\begin{center}
\includegraphics[scale=0.6]{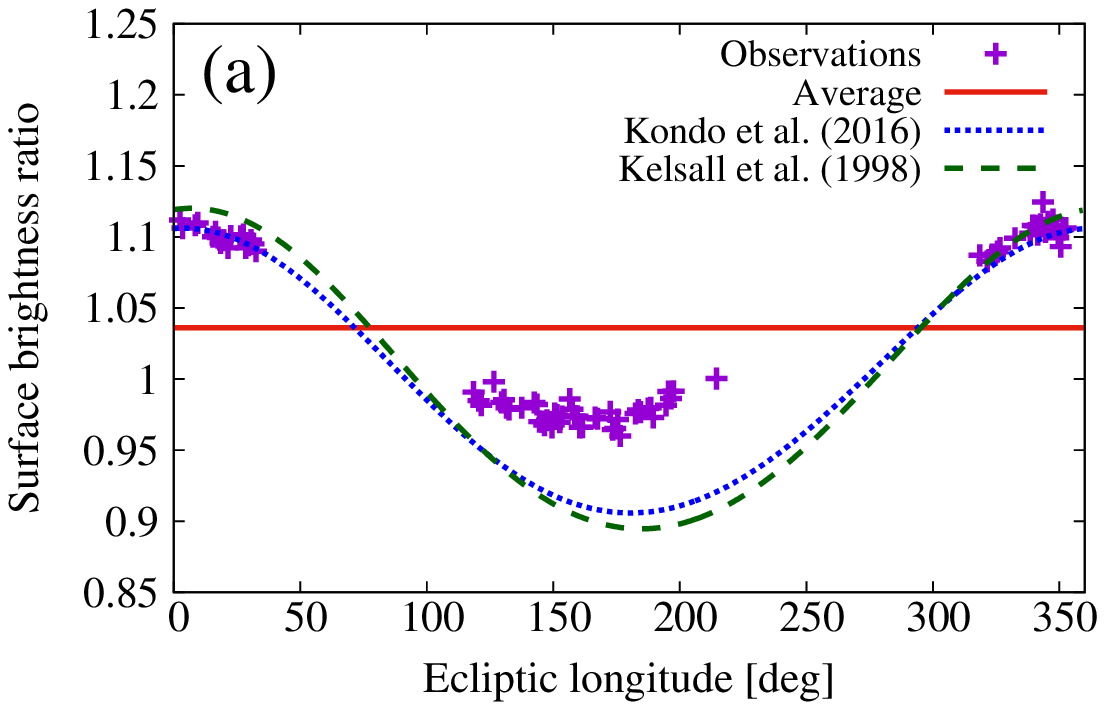}
\label{fig:ratio9}
\includegraphics[scale=0.6]{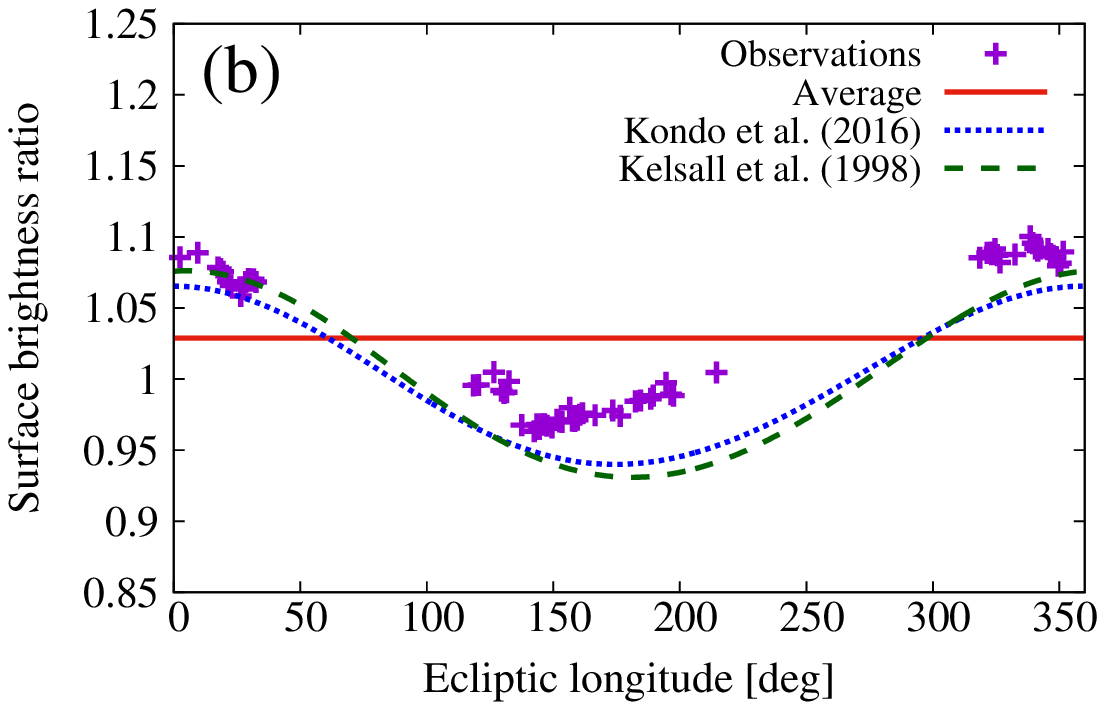}
\label{fig:ratio18}
\caption{
Surface brightness ratio of the trailing to leading observations as a function of the ecliptic longitude in the (a) $9\,\micron$ and (b) $18\,\micron$ bands.
Here, we compare the leading and trailing observations headed in the same direction, 
which means that the ecliptic longitude of the Earth for the leading observation differs from that for the trailing observation by $180^{\circ}$ (see Figure \ref{fig:fig}).
The red solid lines show the mean value of observations, 1.036 for $9\,\micron$ and 1.028 for $18\,\micron$.
The blue dotted and green dashed lines show the analytical model based on the smooth cloud model provided by \citet{Kondo2016} and \citet{Kelsall1998}, respectively.
}
\label{fig:flux-ratio1}
\end{center}
\end{figure}

We analyze the all-sky survey data observed by AKARI in the $9\,\micron$ and $18\,\micron$ bands.
The observational data were corrected for reset anomaly and non-linearity of the detector \citep{Ishihara2010} and for effects of cosmic ray \citep{Mouri2011}.
Figure \ref{fig:akari_map} shows the all-sky survey maps observed by AKARI.
The regions that are strongly affected by the Galaxy and the Moon are masked (the region colored with black in Figure \ref{fig:akari_map}; see also \citealt{Kondo2016}).
Left and right panels show the surface brightnesses observed in the leading and trailing directions of the Earth, respectively.
The data are provided at intervals of $1^{\circ}$ for the ecliptic longitude and latitude, respectively.
In Figure \ref{fig:akari_map}, observed infrared sky brightness at low ecliptic latitude from $-30^{\circ}$ to $30^{\circ}$ is much greater than that at high latitude  and there are differences in the low-latitude brightnesses between the leading and trailing observations.
In order to quantitatively evaluate the difference in the surface brightness between leading and trailing directions, 
we compare the brightnesses integrated with respect to ecliptic latitude on the interval from $-30^{\circ}$ to $30^{\circ}$.
In addition, we only use data with no masked regions for each ecliptic longitude.
In consequence, we obtain about 30--70 integrated brightnesses as a function of ecliptic longitude.
Figure \ref{fig:flux-ratio1} shows the surface brightness ratio of the trailing to leading observations as a function of ecliptic longitude, where we compare the leading and trailing observations headed in the same direction, which means that the ecliptic longitude of the Earth for the leading observation differs from that for the trailing observation by $180^{\circ}$ (see Figure \ref{fig:fig}).
For comparison, we also plot the brightness ratio derived from a simple smooth cloud model proposed by \citet{Kelsall1998} and \citet{Kondo2016}.
In the model, the profiles of temperature $T$ and density $\rho$ are assumed as
\begin{eqnarray}
 T=T_{0} \left( \frac{R}{1\,{\rm AU}} \right)^{-p_c},
\end{eqnarray}
\begin{eqnarray}
 \rho=\rho_{0} \left( \frac{R_{c}}{1\,{\rm AU}}\right)^{-q_c} \exp{\left( -C_{1}g^{C_{2}} \right)}, \label{eq:rho}
\end{eqnarray}
where
\[
 g = \left\{ \begin{array}{ll}
    \frac{(Z_{c}/R_{c})^{2}}{2C_{3}} & (Z_{c}/R_{c}<C_{3}), \\
    Z_{c}/R_{c}-C_{3}/2 & (Z_{c}/R_{c} \geq C_{3}),
  \end{array} \right.
\]
$R$ is the distance from the Sun, $R_{c}$ and $Z_{c}$ are, respectively, the distance from the center of the zodiacal cloud and the vertical distance from the mid-plane of zodiacal cloud, and $T_{0}$, $\rho_{0}$, $p_{c}$, $q_{c}$, $C_{1}$, $C_{2}$, and $C_{3}$ are parameters determined from fitting.
We use the offset of the cloud center, $T_{0}$, $\rho_{0}$, $p_c$, $q_c$, $C_{1}$, $C_{2}$, and $C_{3}$ derived by \citet{Kondo2016} and \citet{Kelsall1998}.
We numerically calculate the integrated surface brightness from
\begin{eqnarray}
 F_{\nu}= \int_{0}^{\infty} \int_{-30^{\circ}}^{30^{\circ}} \frac{\kappa \rho B_{\nu}(T)}{4\pi r^{2}} r\Delta \theta dr d\phi,
\end{eqnarray}
where  $r$ is a distance from the Earth, $\Delta \theta$ is a viewing angle in direction of ecliptic longitude, $\phi$ is the ecliptic latitude and $\kappa$ is the opacity.
We choose a constant $\kappa$ for simplicity. 

Figure \ref{fig:flux-ratio1} shows that there is a large variation in the observational brightness ratio with the amplitude of about 6--7\% and the mean value is about $1.03$.
In contrast, the mean value of the analytical model is nearly unity.
The deviation from the mean value, $\pm 0.06$, is mainly caused by the orbital eccentricity of the Earth orbit (inclination also affects the ratio, but this effect is much smaller than that caused by eccentricity).
The leading and trailing brightnesses of the same ecliptic longitude were measured from the Earth located at the different positions, 
which means that the ecliptic longitude of the Earth for the leading observation differs from that for the trailing observation by $180^{\circ}$.
Hence, due to the eccentricity of the Earth, the difference in the distances from the Sun is up to about 3\% (i.e. $2e_{\oplus}a_{\oplus} \simeq 0.033\,{\rm AU}$, $e_{\oplus}=0.0167$ and $a_{\oplus}=1\,{\rm AU}$ are orbital eccentricity and semi major axis of the Earth) and it leads to the difference in temperature and spatial number density of surrounding dust particles. 
In addition, the leading-trailing asymmetry in the local dust number densities caused by circumsolar ring component also affects the surface brightness ratio as we discuss below.
This ring component enhances the trailing brightness compared to the leading brightness.
Because we apply a simple smooth density distribution (Equation \ref{eq:rho}), the effect of the ring component is not included in the analytical model.

\begin{figure}[h]
\begin{center}
\includegraphics[scale=0.6]{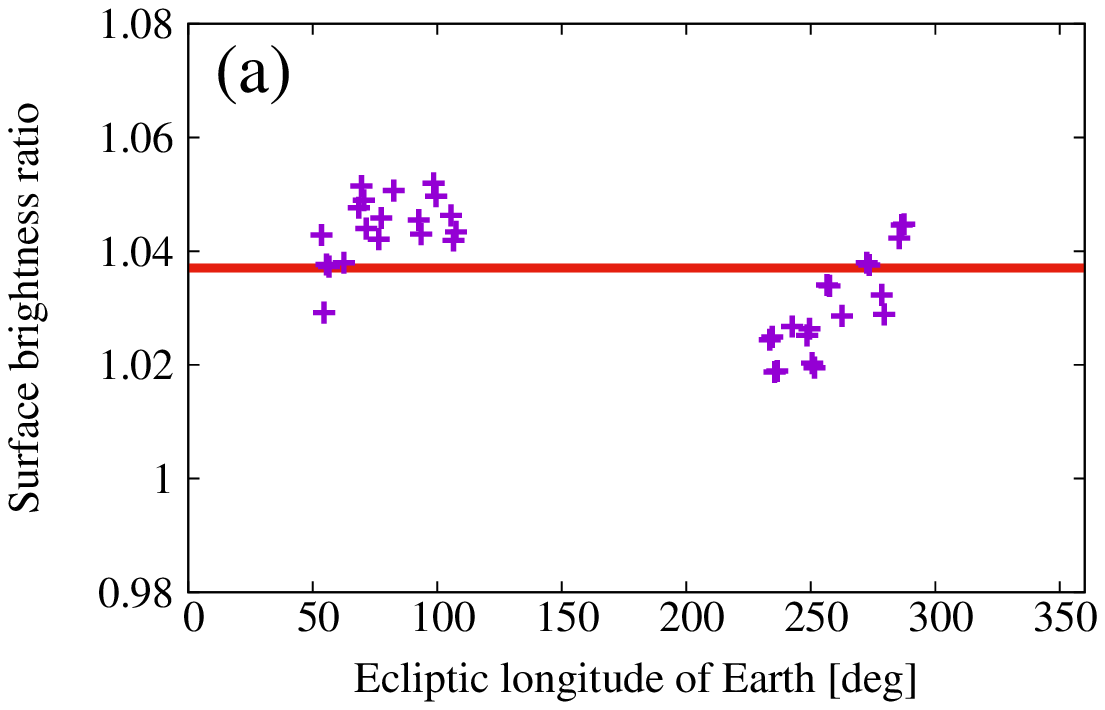}
\label{fig:ratio9}
\includegraphics[scale=0.6]{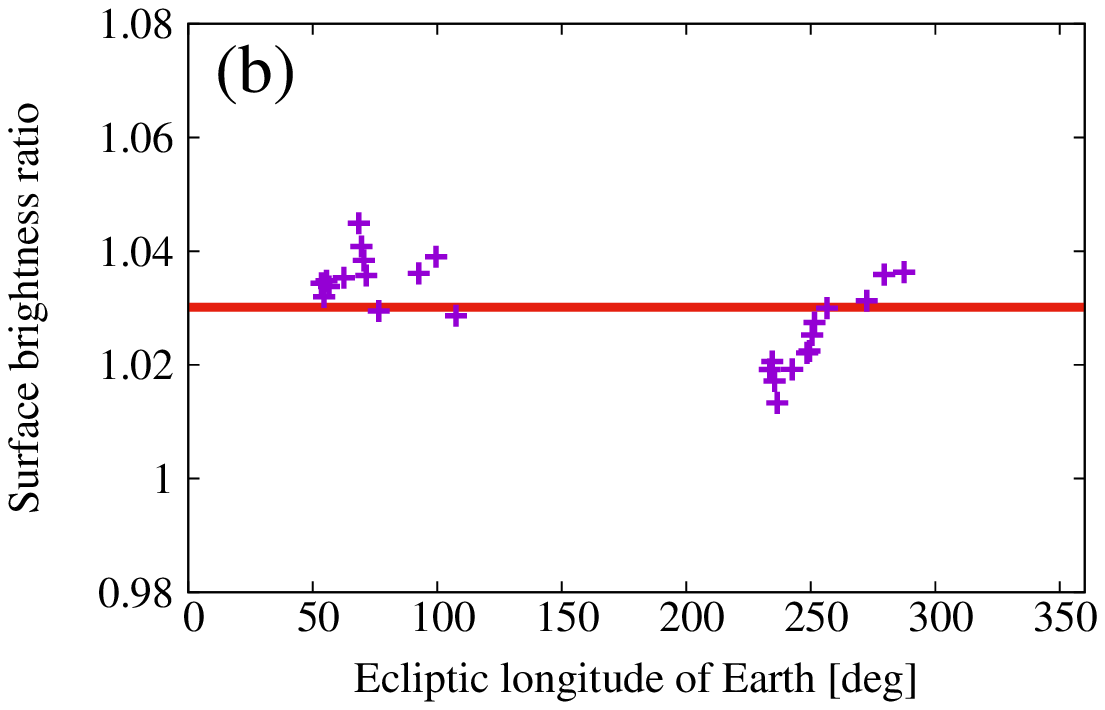}
\label{fig:ratio18}
\caption{
Surface brightness ratio of the trailing to leading observations as a function of the ecliptic longitude of Earth in the (a) $9\,\micron$ and (b) $18\,\micron$ bands.
Here, we compare the leading and trailing observations headed in the opposite directions, which means that the ecliptic longitudes of the Earth are the same (see Figure \ref{fig:fig}).
The red solid lines show the mean value of the observations, 1.037 for $9\,\micron$ and 1.030 for $18\,\micron$.
The standard deviations of the data sets are 0.01 for $9\,\micron$ and 0.008 for $18\,\micron$.
}
\label{fig:flux-ratio2}
\end{center}
\end{figure}

Figure \ref{fig:flux-ratio2} shows the surface brightness ratio of the trailing to leading observations as a function of the ecliptic longitude of the Earth.
In contrast to Figure \ref{fig:flux-ratio1}, the brightness ratio changes insignificantly because the effect of the eccentricity of Earth orbit is excluded.
However, even though brightnesses were measured at the same distance from the Sun, the brightness in the trailing direction is always greater than that in the leading direction.
The mean value of the ratio remains 1.037 in the $9\,\micron$ band and 1.030 in the $18\,\micron$ band.
These values are similar to the measurement by IRAS \citep{Dermott1994}.
This mean difference in brightness is considered to come from the leading-trailing asymmetry in the local dust number density caused by the circumsolar ring.
Although  seasonal variations seem to exist even without the effect of the eccentricity, we use the mean value to make a comparison with the theoretical results shown in following sections.

\section{Model for Orbital Evolution of Dust Particles}
\label{sec:model}

\subsection{Basic Equations}
We numerically solve the three dimensional restricted three-body
problem of the Sun, the Earth, and a dust particle, including radiation
force from the Sun by orbital integration with the fourth-order Hermite scheme \citep{Makino}.
The basic equation for the orbital motion of a dust
particle of mass $m$ and radius $s$ with position at $\bm{R}$ and velocity $\bm{V}$ 
in the heliocentric coordinate system is given by (e.g., \citealt{Burns1979})
\begin{align}
	\frac{d^2 \bm{R}}{dt^2}=&-GM_{\odot}(1-\beta)\frac{\bm{R}}{|\bm{R}|^3}-GM_{\oplus}\left( \frac{\bm{R}-\bm{R}_{\oplus}}{|\bm{R}-\bm{R}_{\oplus}|^3}+\frac{\bm{R}_{\oplus}}{|\bm{R}_{\oplus}|^3}\right) \nonumber \\
	&-\frac{GM_{\odot}\beta_{\rm PR}}{c|\bm{R}|^2}\left( \frac{\bm{V}\cdot\bm{R}}{R}\frac{\bm{R}}{R}+\bm{V} \right),  \label{eq:eom}
\end{align}
where $M_\sun$ is the solar mass, $M_{\oplus}$ and $\bm{R}_{\oplus}$ are
the mass and position vector of the Earth, respectively, $\beta$ and
$\beta_{\rm PR}$ are dimensionless parameters related to the radiation
pressure and the P-R drag, respectively, $G$ is the gravitational constant, and $c$ is
the speed of light.  The dimensionless parameter $\beta$, the ratio of
solar radiation pressure to solar gravity, is defined as
\begin{align}
	\beta &\equiv \frac{SAQ_{\rm pr}/c}{GM_{\odot}m/R^{2}} \nonumber \\
	&\simeq \left( \frac{s}{0.285\,\micron}\right)^{-1} \left( \frac{\rho_{\rm s}}{2 \ \rm{g\,cm^{-3}}}\right)^{-1} \left( \frac{Q_{\rm pr}}{1}\right), \label{eq:beta}
\end{align}
where $S\equiv L_{\odot}/4\pi R^{2}$ is the energy flux density of Sun,
$L_\sun$ is the solar luminosity, $A\equiv \pi s^{2}$ is the geometrical cross section of the grain, $Q_{\rm pr}$ is the radiation pressure efficiency factor of the dust grain. 
The first and second terms on the right hand side in Equation (\ref{eq:eom}) are
the gravity of the Sun and the Earth, respectively, though the
factor $1-\beta$ in the first term comes from the addition of solar radiation pressure. 
Due to the radiation pressure, particles feel that
the Sun is less massive.  The third term on the right hand side
represents the P-R drag force due to the solar wind as well as the solar radiation. 
The ratio of the P-R drag force due to the solar wind to the P-R drag force due to the stellar radiation is about 30\% (e.g., \citealt{Minato2004}), so that we set $\beta_{\rm PR}=1.3\beta$ .  
Note that in our calculations, the perturbations from planets other than the Earth are not included.  
This is because we
need to conduct a number of calculations with various values of $\beta$
and the full-body calculation with small $\beta$ value is very
time-consuming.  
The effect of perturbations from planets especially Jupiter is discussed in Section \ref{sec:discuss}.

\subsection{Initial Conditions}
We assume that the particles are originated from main-belt asteroids or Jupiter-family comets.
Jupiter-family comets rotate around the Sun with relatively short periods and are considered to be main parent comets of IDPs (e.g., \citealt{Nesvorny2010}). 
Parent bodies feel negligible radiation pressure because of large masses, whereas ejected dust particles are affected by radiation pressure according to the value of $\beta$.
The orbital elements of dust particles just after their release from parent bodies are related to those of parent bodies and particle's $\beta$ (e.g., \citealt{Strubbe2006}, \citealt{Kobayashi2008}).
If a particle with $\beta$ is ejected from a parent body at the distance $R$ from the Sun, the initial eccentricity and semi major axis of the particle are given by
\begin{eqnarray}
	a_{\rm ini}=\frac{1-\beta}{1-2\beta a_{\rm p}/R}a_{\rm p},
\label{eq:aini}
\end{eqnarray}
\begin{eqnarray}
	e_{\rm ini}=\sqrt{ \frac{ \left( 2a_{\rm p}/R -1 \right)\beta - \left( 2\beta a_{\rm p}/R -1 \right)e_{\rm p}^{2} }{1-\beta} },
\label{eq:eini}
\end{eqnarray}
where $a_{\rm p}$ and $e_{\rm p}$ are the semi major axis and eccentricity of the parent body, respectively.
Here, we assumed that the ejection speed relative to the parent body is much smaller than the orbital velocity of the parent body.
Equation (\ref{eq:eini}) indicates that if the parent body has a circular orbit, the orbital eccentricity of the produced particle increases with $\beta$ and exceeds unity for $\beta > 0.5$.
We set the orbital elements of the parent bodies according to the data provided at JPL small-body Database Search Engine (\url{http://ssd.jpl.nasa.gov/sbdb_query.cgi#x}).
The average orbital eccentricity and inclination of the main-belt asteroids are $e_{\rm a,ave} = 0.15$ and $i_{\rm a,ave}= 0.13$ radian, respectively.
The average eccentricity and inclination of Jupiter-family comets are $e_{\rm c,ave}=0.54$ and $i_{\rm c,ave}=0.24$ radian, which are 3.6 and 1.8 times larger than those of main-belt astroids, respectively.
High temperature makes comets active so that cometary dust particles are set to be ejected when the parent comets are located at $r<1.5\,{\rm AU}$.
In order to save the computational cost, if the distance of the perihelion of particle's orbit to the Sun, $a(1-e)$, is larger than $1.6\,{\rm AU}$ (which corresponds to the 2:1 MMR of the Earth), the orbital evolution is given without planetary perturbation, and then the orbits of particles are determined by \citep{Wyatt1950}
\begin{eqnarray}
	a(1-e^{2})e^{-4/5}={\rm const}.
	\label{eq:const}
\end{eqnarray}
This simplification does not affect the result because our preliminary calculations suggest that almost no dust particle is trapped in MMRs beyond the 2:1 MMR.

Note that cometary dust may include icy components. Although the size evolution due to sublimation
affects orbital evolution (\citealt{Kobayashi2009}, \citealt{Kobayashi2011}), ice sublimation is ineffective inside 5\,AU (\citealt{Kobayashi2010}). 
Therefore, we do not consider the ice sublimation of drifting dust.

\section{Orbital Evolution of Dust particles}
\label{sec:orbital}

\subsection{Typical Orbital Evolutions}
Here, we briefly summarize the orbital evolution of the dust particles under the radiation from the Sun and the gravity from the Sun and the Earth.
We also explain why the trapped particles produce a leading-trailing density asymmetry around the Earth.

In the course of radial drift due to P-R drag, some particles are trapped in MMRs with the Earth. 
Although MMRs reside at both of interior and exterior regions of Earth's orbit, 
we focus on the exterior resonances because AKARI observed only the outer region (see Figure \ref{fig:fig}).  
The location of the $p:p+j$ MMR of the Earth's orbit can be written as
\begin{eqnarray}
a_{p,j}=(1-\beta)^{1/3}\left( \frac{p+j}{p} \right)^{2/3} \ {\rm AU}. \label{eq:mmr}
\end{eqnarray}
The conjunction of the Earth and trapped particles occurs at the same configuration. 
Since trapped particles mainly have larger eccentricities than that of the Earth, 
the orbital phases of trapped particles determine the conjunction distances between the Earth and the trapped particles. 
If the conjunction occurs at the perihelion of the particle, 
the particle is strongly perturbed by the Earth due to short distance from the Earth and the orbit of the particle becomes unstable. 
In contrast, if the conjunction occurs at the particle's aphelion,
the particle is weakly perturbed by the Earth and can stay in the orbit for a long term.
\begin{figure}[h]
\begin{center}
\includegraphics[scale=0.6]{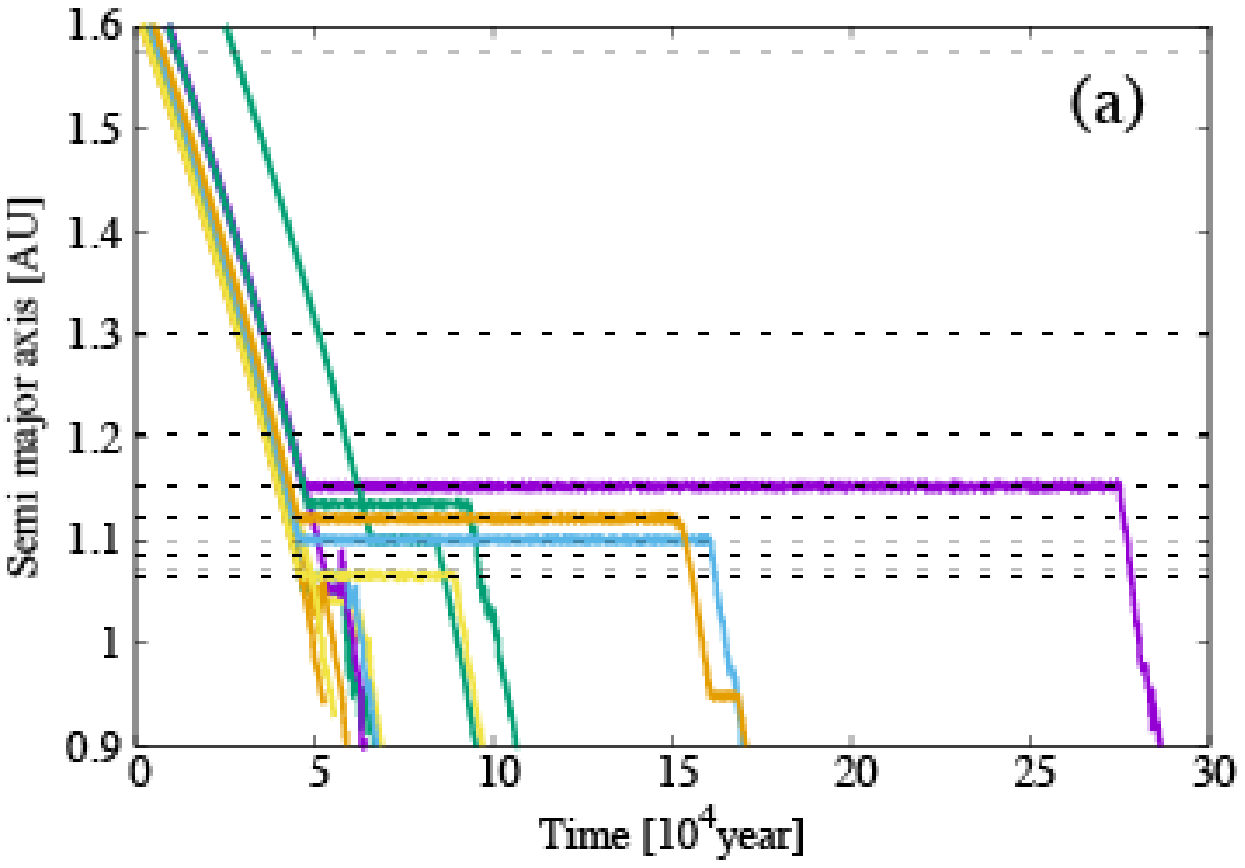}
\label{fig:semi} \includegraphics[scale=0.6]{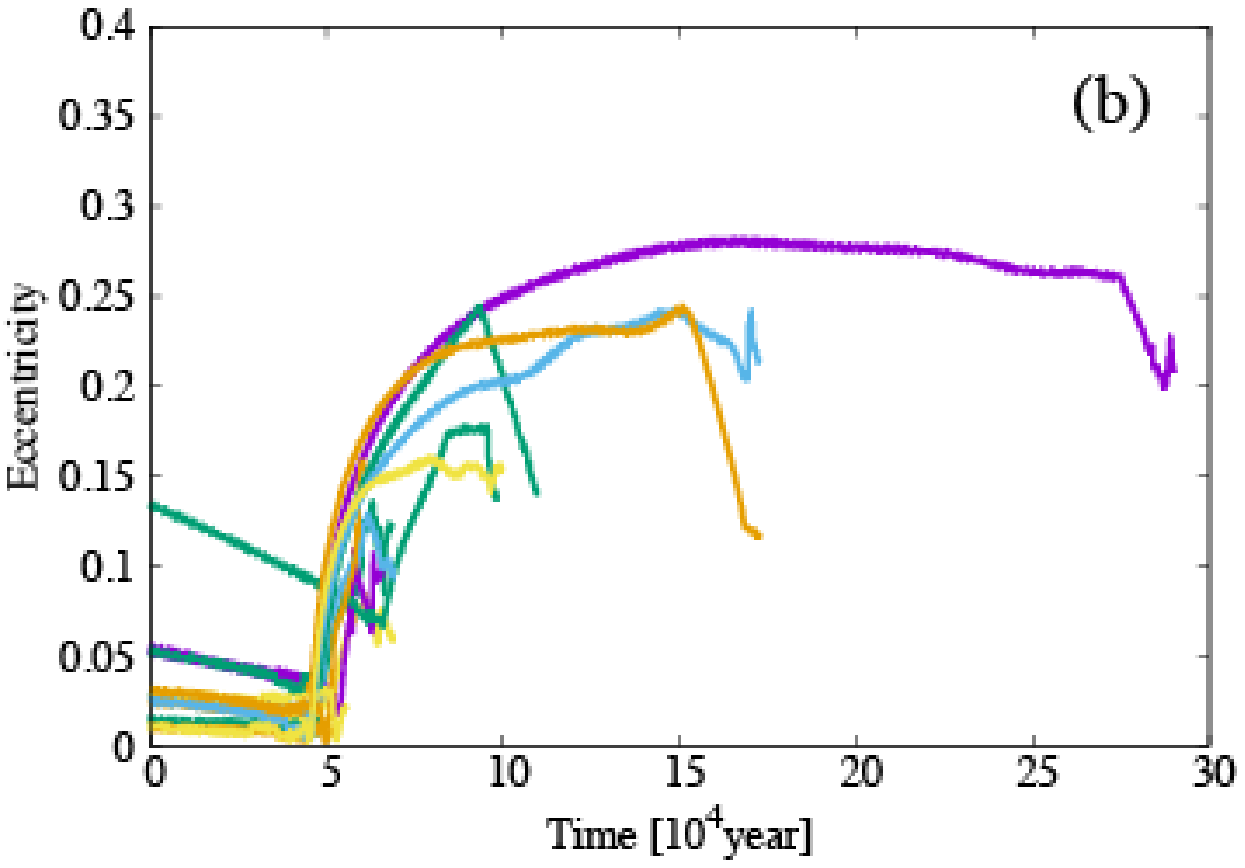}
\label{fig:ecc} \caption{Typical orbital evolutions of asteroidal dust
particles with $\beta=0.02$.  The locations of the $p:p+1$ resonances
from $p=1$ to $p=9$ are denoted by black dashed lines from top to
bottom.  } \label{fig:typical}
\end{center}
\end{figure}
Figure \ref{fig:typical} shows typical orbital evolutions of asteroidal particles with $\beta=0.02$.  
Some particles are trapped in MMRs of the Earth for longer than $5 \times 10^{4}$ years, 
so that the trapped particles are in stable phases. 
However, the trapped particles gradually increase their orbital
eccentricities and finally get out of the resonances due to close encounters with the Earth. 

\begin{figure*}[h]
\begin{center}
\includegraphics[scale=0.45]{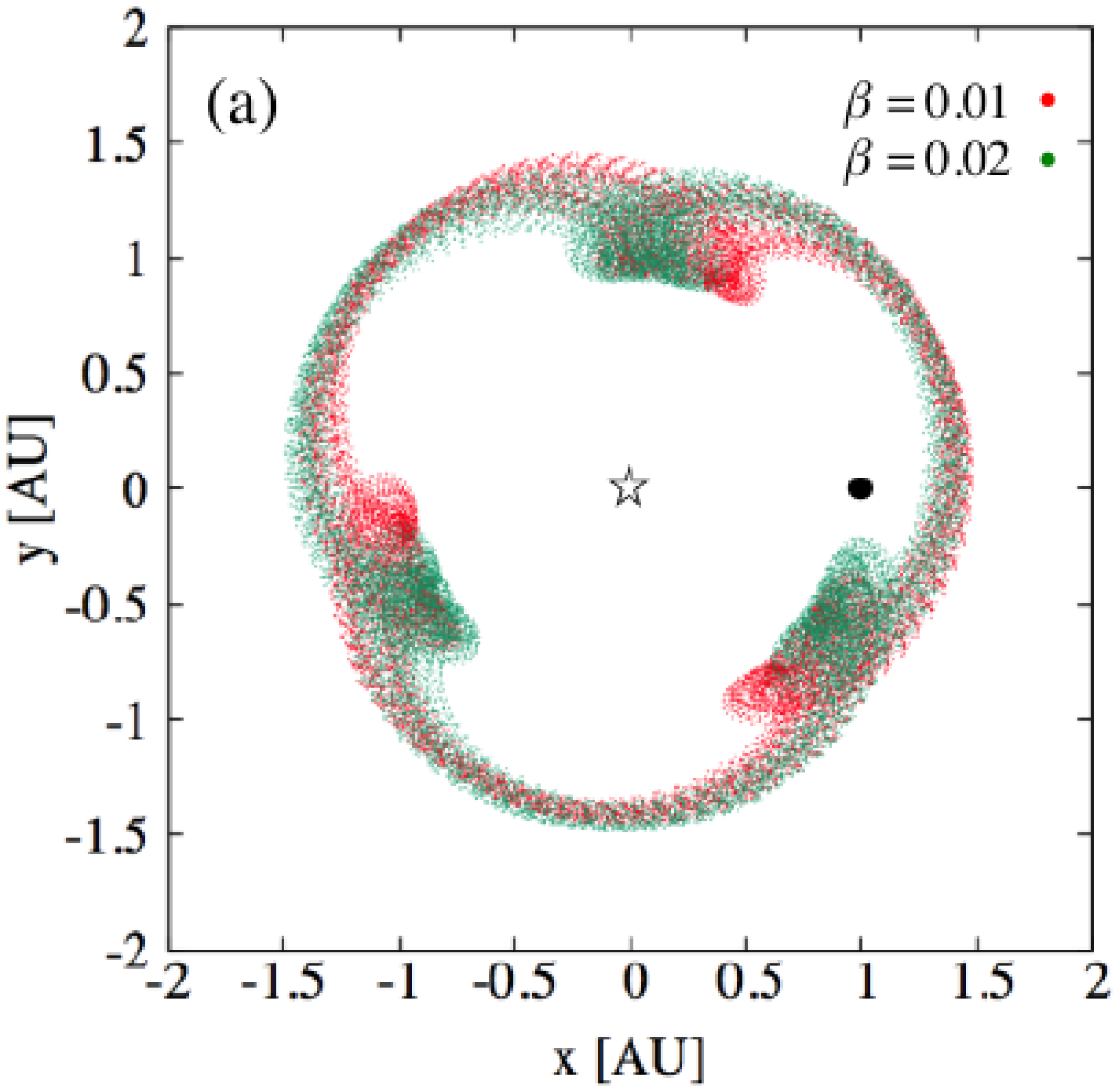}
\label{fig:3mmr} 
\includegraphics[scale=0.45]{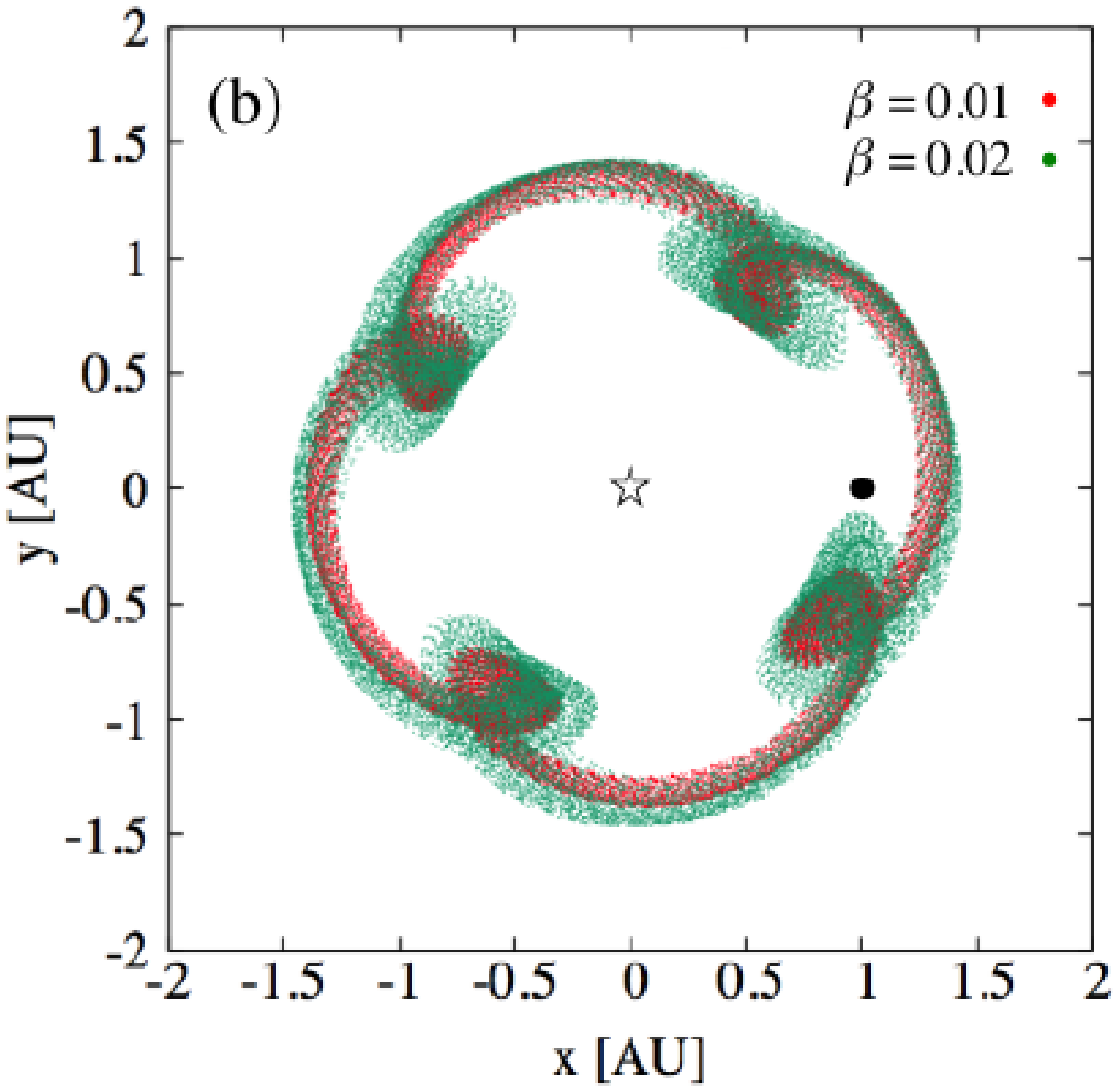}
\label{fig:4mmr}
\\
\includegraphics[scale=0.45]{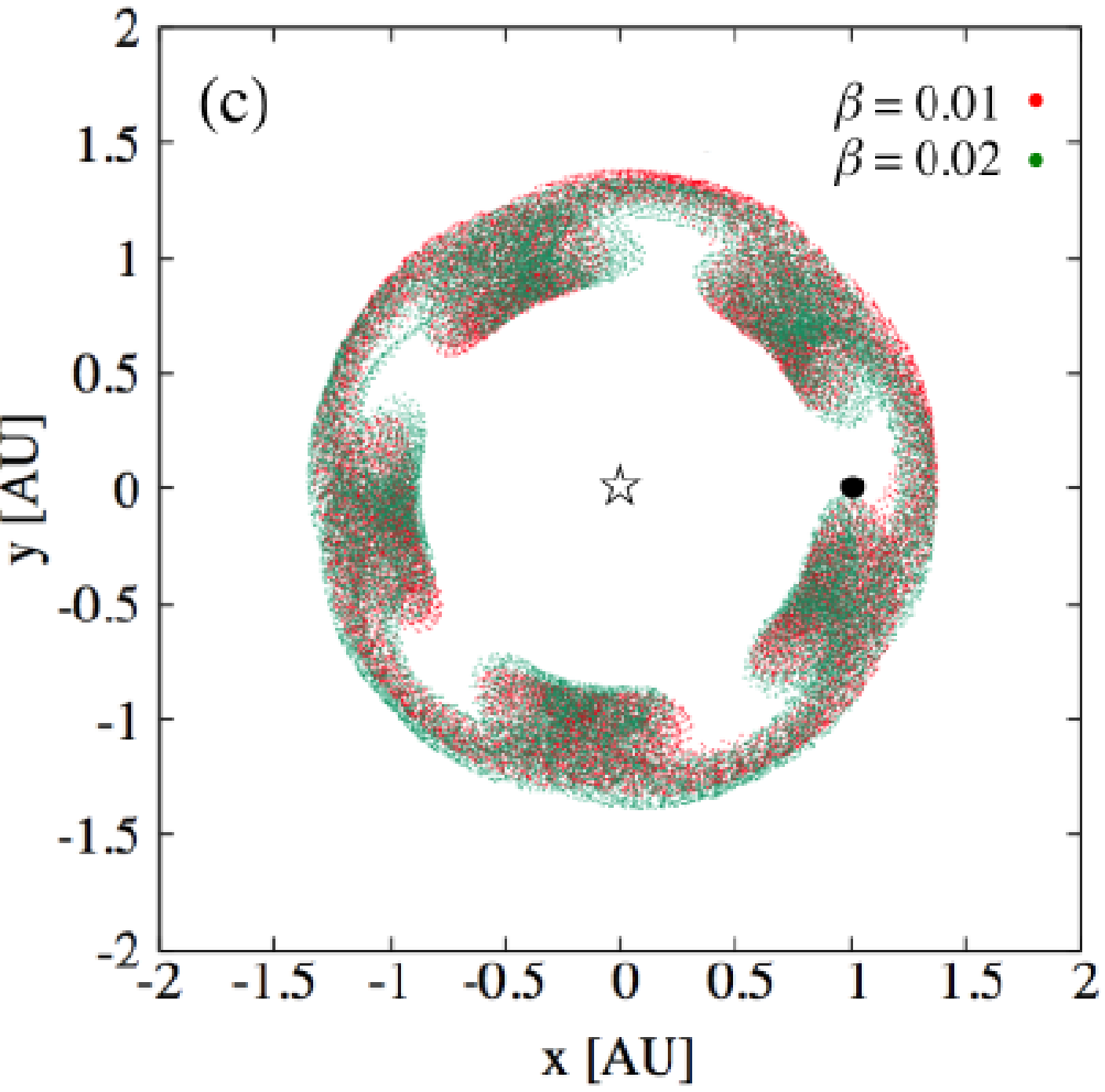}
\label{fig:5mmr}
\includegraphics[scale=0.45]{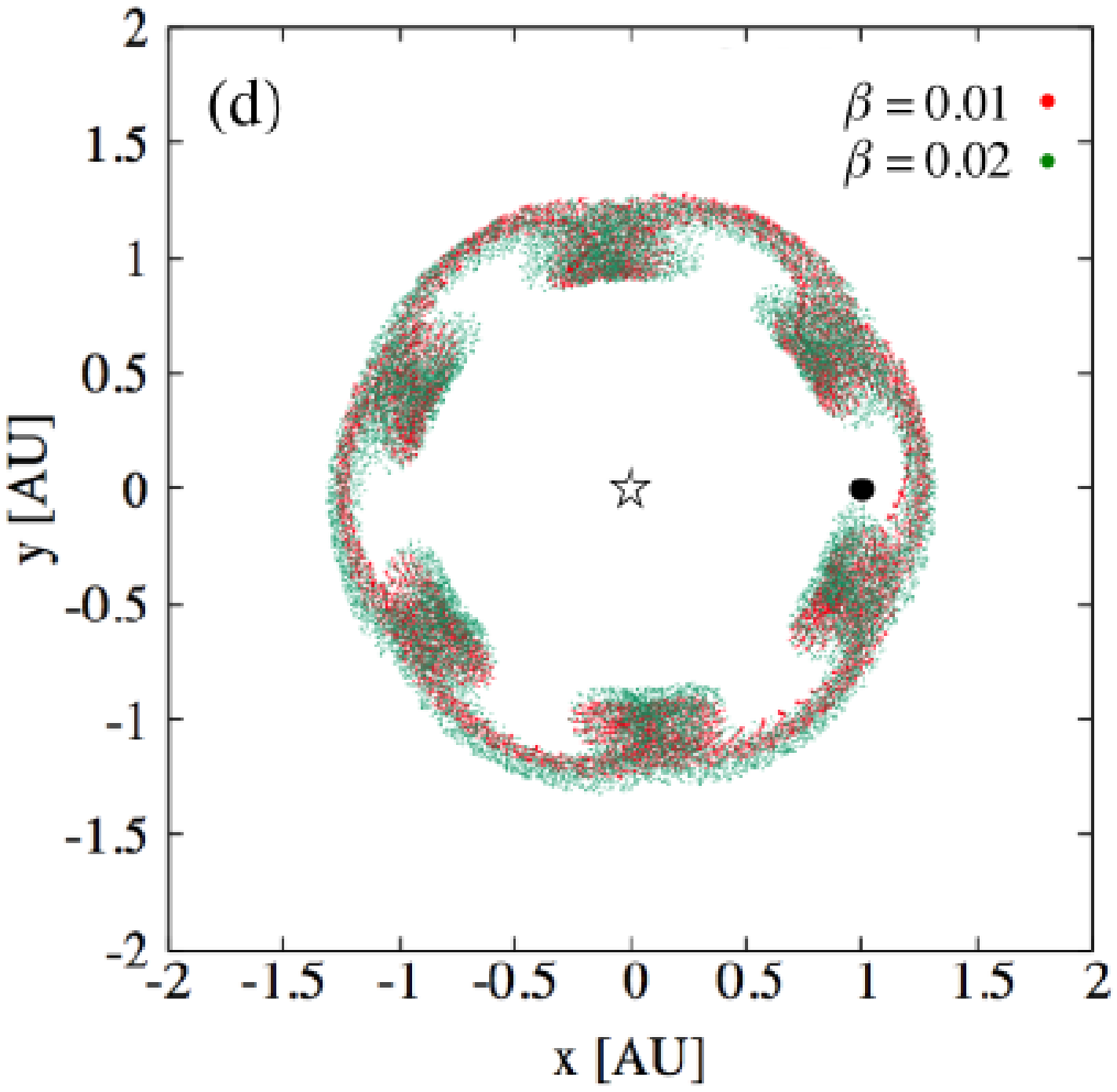}
\label{fig:6mmr}
\caption{
Example paths of an asteroidal particle with $\beta=0.01$ and $0.02$ trapped in $p:p+1$ MMRs with the Earth in the frame co-rotating with the mean motion of the Earth for (a) $p=3$, (b) $p=4$, (c) $p=5$ and (d) $p=6$.
The positions of the particle along each path are drawn at equal time intervals.
The Sun is located on the origin and the guiding center of the Earth that rotates about the Sun in a circle of $1\,{\rm AU}$ with angular velocity equal to the Earth's mean motion is denoted by a black filled circle.}
\label{fig:mmrs}
\end{center}
\end{figure*}
Figure \ref{fig:mmrs} shows the paths for single particles with $\beta=0.01$ and $0.02$ trapped in $p:p+1$ ($p=3, 4, 5$ and $6$) MMRs with the Earth in the frame co-rotating with the mean motion of the Earth.
The conjunction between the Earth and the trapped particle happens around the aphelion of a particle's orbit.
The paths of the particles have ``loops'', which correspond to the perihelia of the orbits of trapped particles.
A particle trapped in a $p:p+1$ resonance has $p$ ``loops''.
The particles stay in ``loops'' for a longer time than the other positions of orbits, so that the number density distribution of the particles is clumpy (e.g., \citealt{Wyatt2003}).
As shown in Figure  \ref{fig:mmrs}, the trailing clumps formed by the ``loops'' are closer to the Earth than leading clumps.

The displacement of the clumps can be understood by considering the change in azimuthal velocity $v_{\theta}$ of a dust particle.
If we ignore solar radiation, the trailing and leading clumps are symmetrically distributed with respect to the Earth. 
If the conjunction occurs at just before the aphelion of dust orbit,
the tangential force experienced by the particle immediately prior to conjunction is smaller than that immediately after
conjunction because of the configuration of the conjunction: 
The dust particle acquires angular momentum from the Earth due to the net result of the
encounter resulting in the increasing in $v_{\theta}$; the next
conjunction becomes closer to or just after aphelion. 
If the conjunction occurs at just after the aphelion of dust orbit, the dust particle loses it's angular momentum due to the encounter resulting in the decreasing in $v_{\theta}$.
Therefore, such encounters keep the conjunction close to
the aphelion; the stable configuration is accomplished (e.g., \citealt{SSD}). 
However, if we consider solar radiation, P-R drag always causes the decreasing in  $v_{\theta}$ for dust particles.
Due to the P-R drag, the conjunction at slightly before the aphelion of
dust orbit becomes stable; larger $\beta$ results in the
stable conjunction farther from the aphelion. 
Therefore, the trailing ``loops'' (perihelia) become closer to the Earth for larger $\beta$. 

\subsection{Spatial Distribution of the Zodiacal Dust Grains}
\label{sec:dist}
In order to investigate the dependence of spatial dust distributions on physical properties ({\rm i.e.} parameter $\beta$),
we conduct orbital calculations of 100--15000 particles for each $\beta$ ranging from $0.0003$ to $0.4$ with different initial conditions.
We divide the calculated area into blocks with azimuthal grid of $1^{\circ}$ and radial grid of $0.005\,{\rm AU}$.
To derive the number density distribution, we count the number of particles in blocks that are stored in orbital data at equal time intervals.
\begin{figure*}[h]
\begin{center}
\includegraphics[scale=0.4]{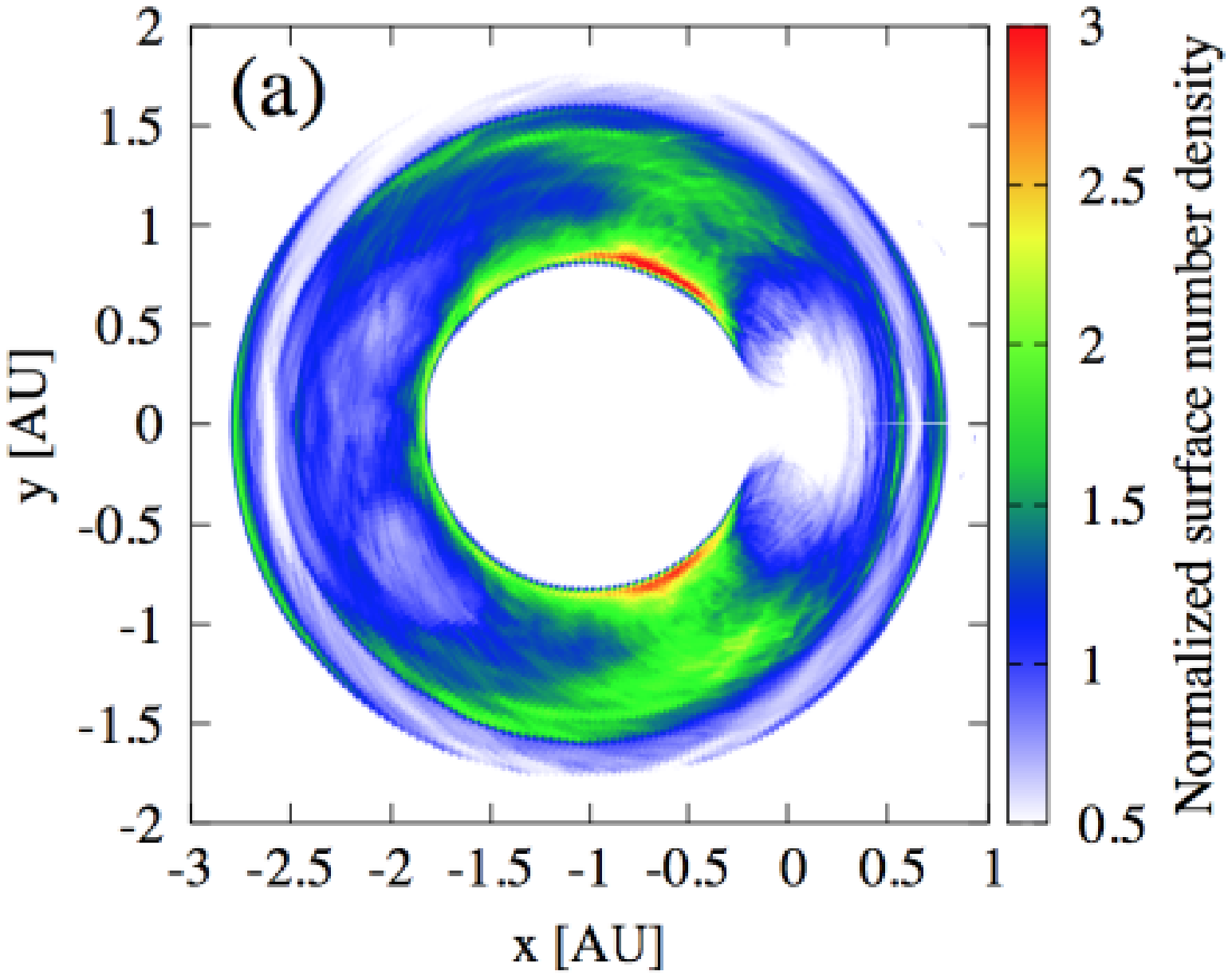}
\label{fig:kinten-map-0.001}
\includegraphics[scale=0.4]{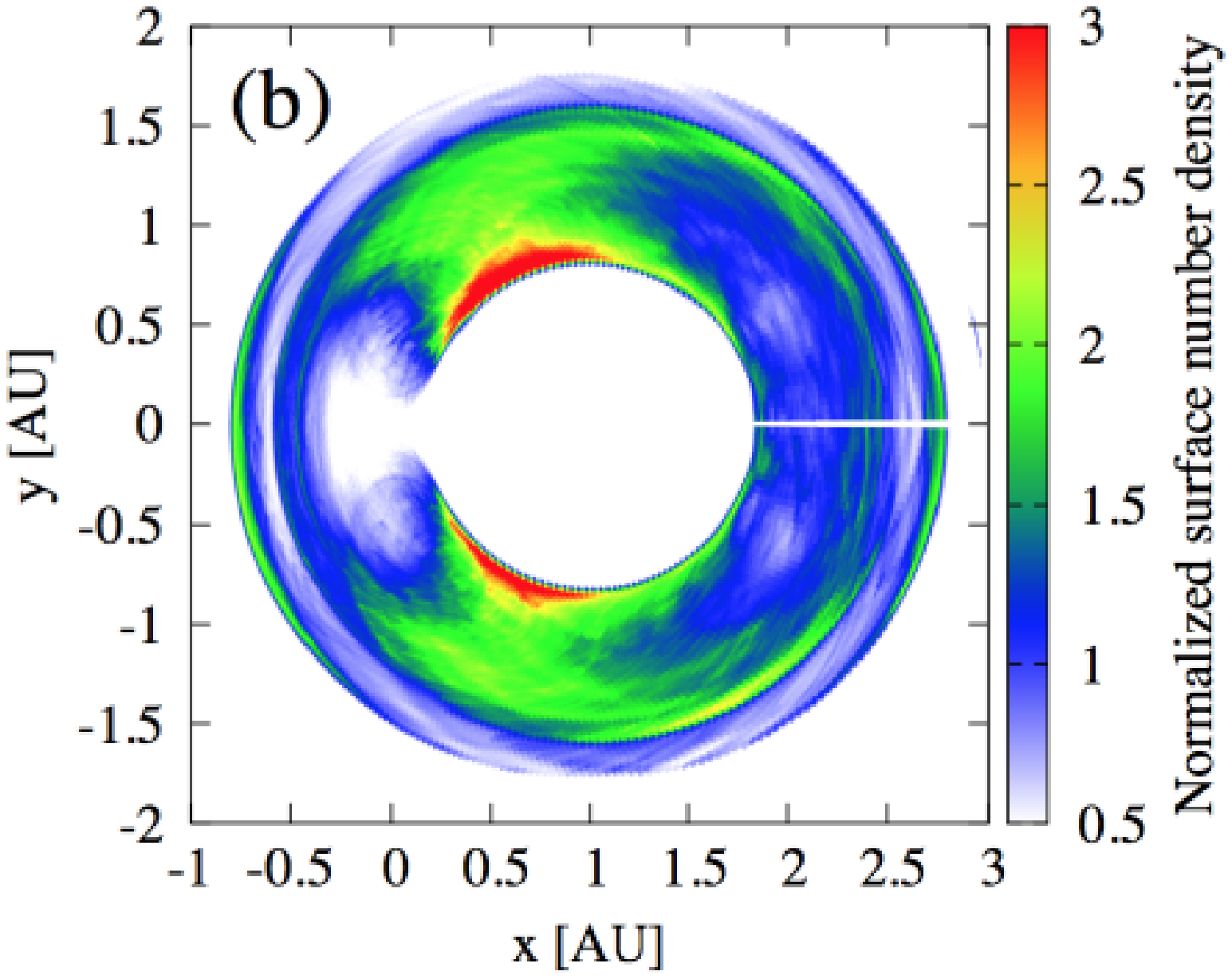}
\label{fig:enten-map-0.001}
\\
\includegraphics[scale=0.4]{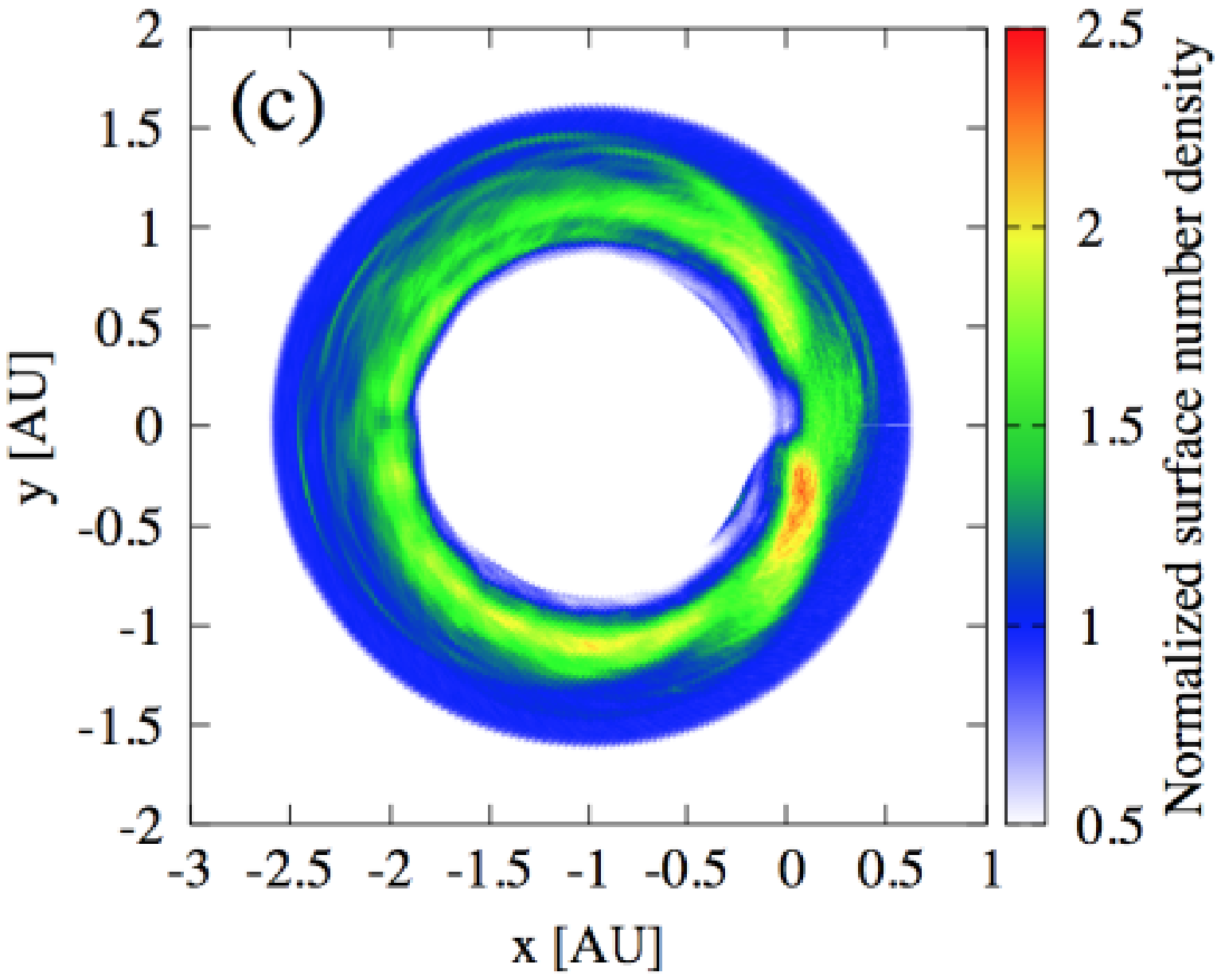}
\label{fig:kinten-map-0.02}
\includegraphics[scale=0.4]{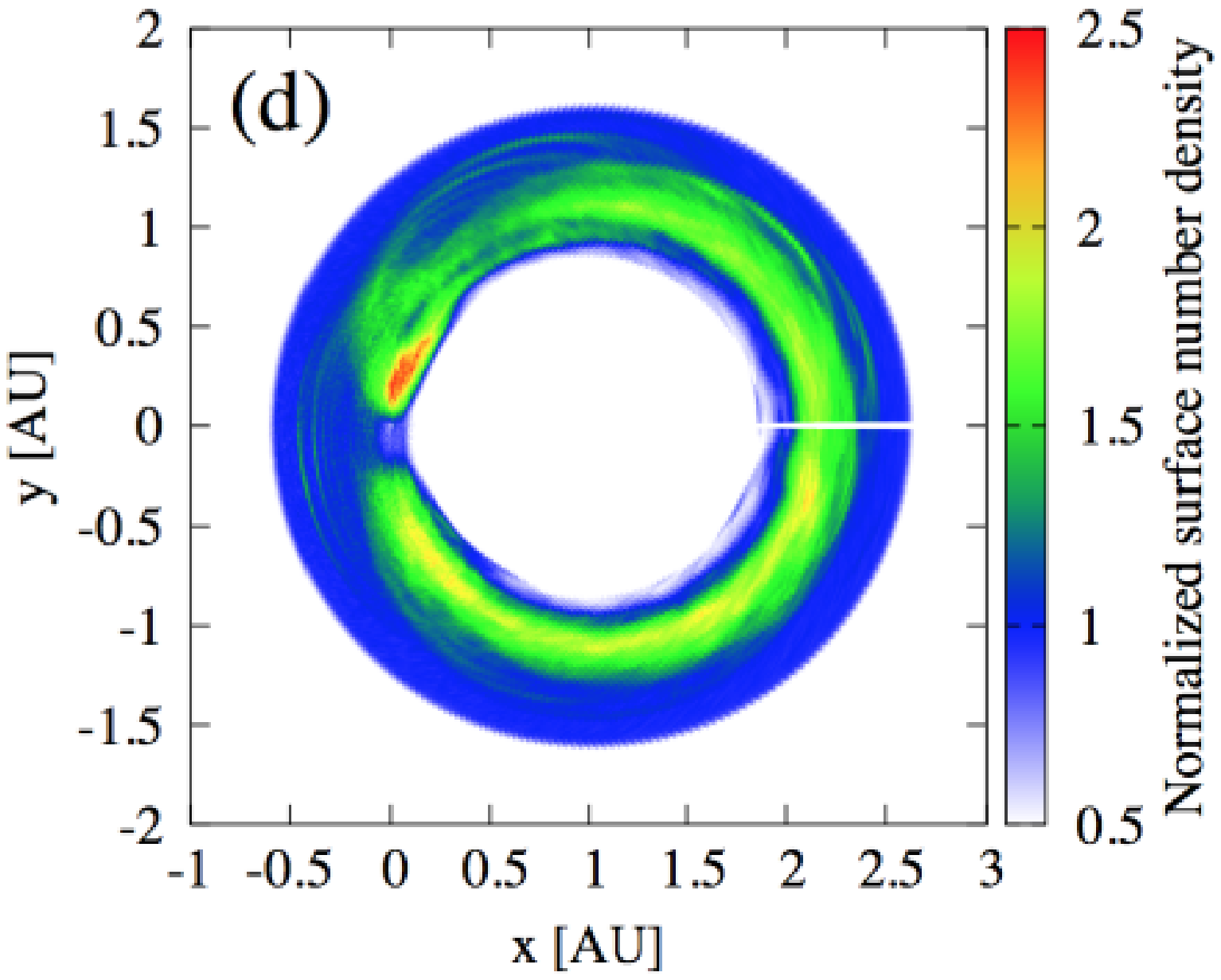}
\label{fig:enten-map-0.02}
\\
\includegraphics[scale=0.4]{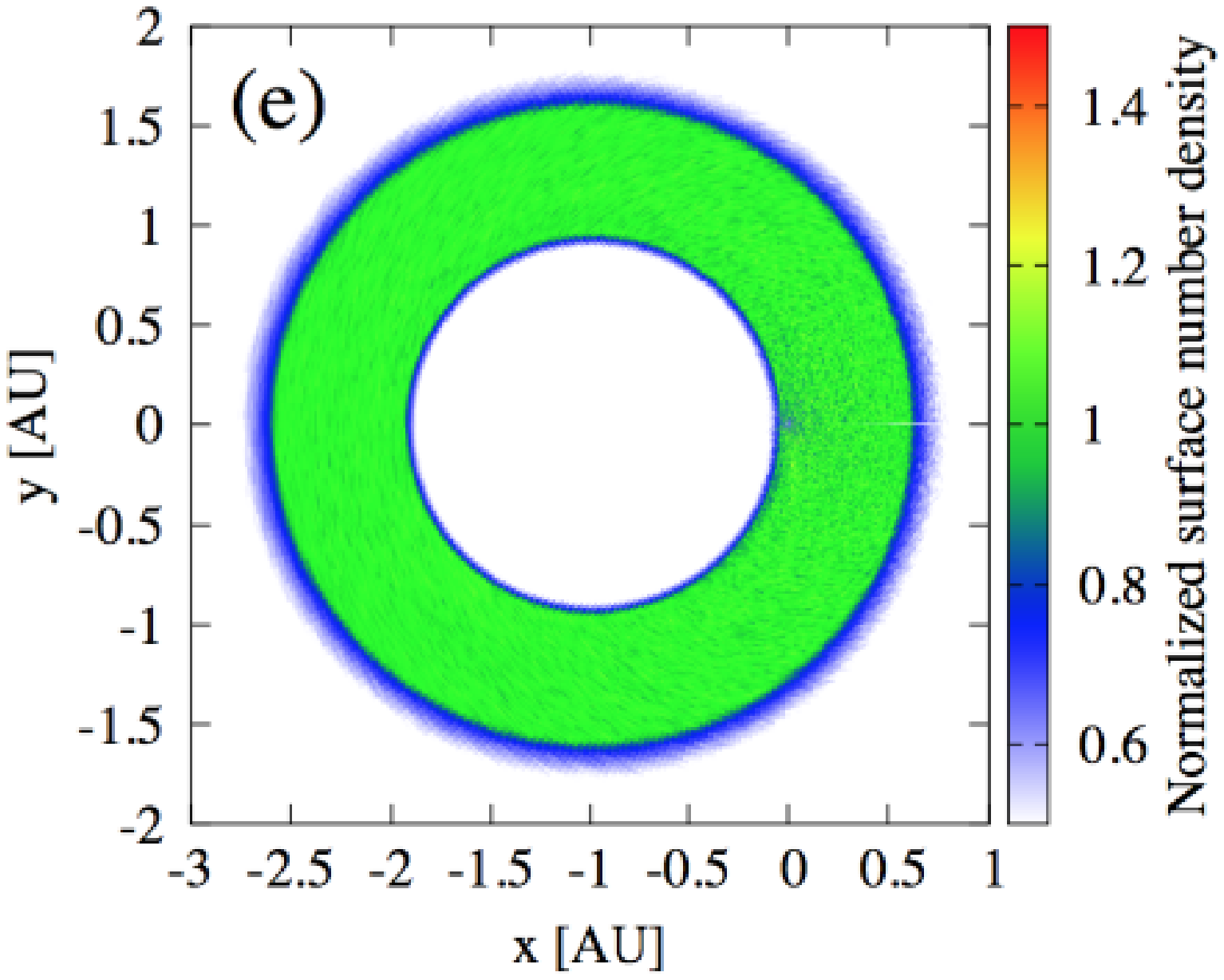}
\label{fig:kinten-map-0.2}
\includegraphics[scale=0.4]{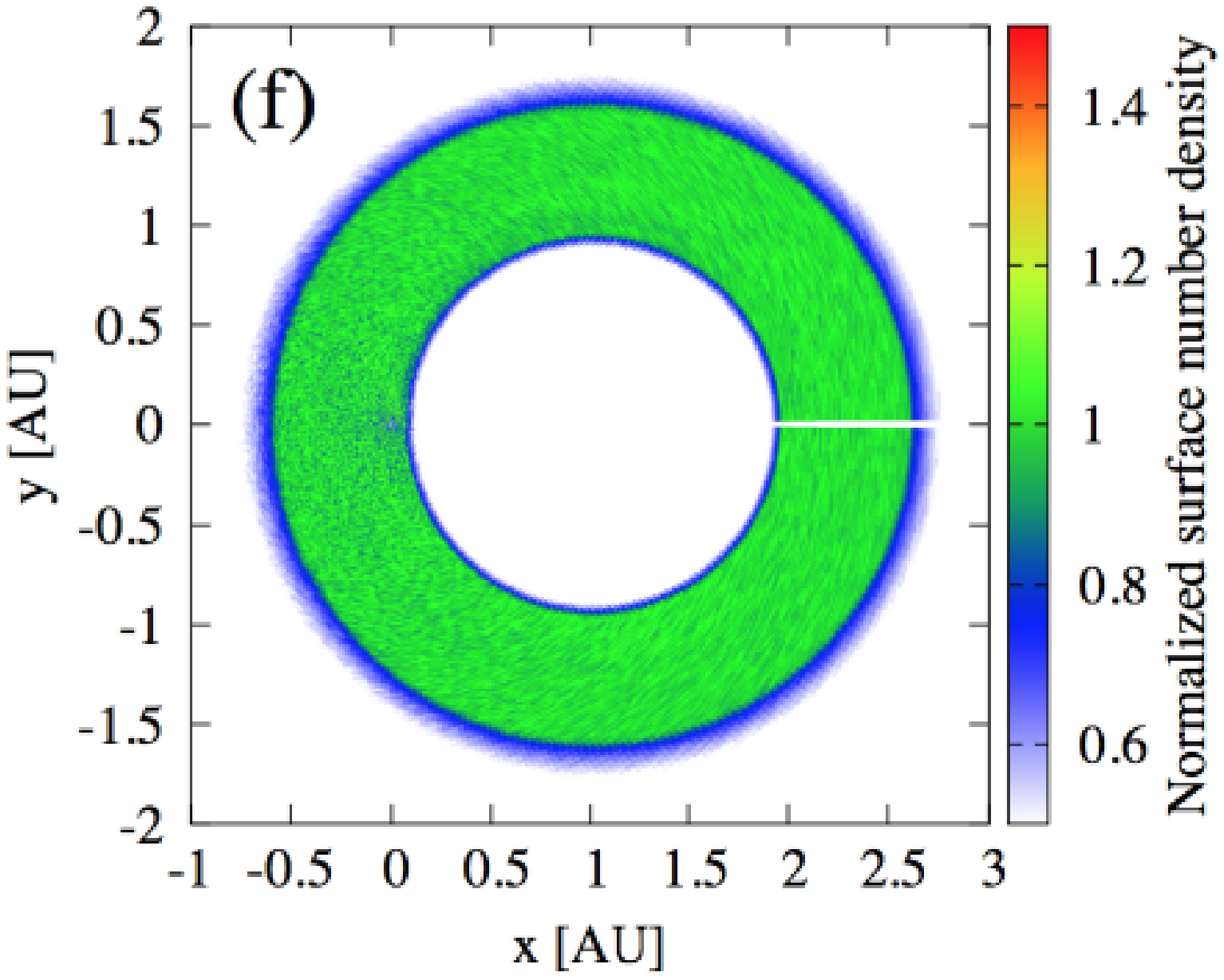}
\label{fig:enten-map-0.2}
\caption{
Normalized surface number density distribution of asteroidal dust particles with $\beta=0.001$ (top panels), $\beta=0.02$ (middle panels), and $\beta=0.2$ (bottom panels).
In the left and right panels, the Earth is located on the perihelion and aphelion of its orbit, respectively.
The Earth rotates counterclockwise.
For intermediate $\beta$ (panels c, d), dust clumps are formed in the trailing directions.
}
\label{fig:dens-map0.02}
\end{center}
\end{figure*}
Figure \ref{fig:dens-map0.02} shows the surface number density $n_{s}$ of asteroidal dust particles with $\beta=0.001$ (top panels), 0.02 (middle panels) and 0.2 (bottom panels).
In the left and right panels, the Earth is located on the perihelion and aphelion of its orbit, respectively.
In each panel, the Earth is located at the origin and rotates in counterclockwise direction, which means that
the leading direction is upward in the left panels and downward in the right panels.
For comparison, we normalize $n_{s}$ by the values at 1.5(2.5)\,${\rm AU}$ from the Sun (Earth), which is slightly smaller than the initial distances from the Sun in orbital integrations.
It can be seen that many particles with $\beta=0.001$ and $0.02$ are trapped in MMRs with the Earth, resulting in the dense region around the Earth.
In contrast, particles with $\beta=0.2$ are smoothly distributed and an asymmetric density distribution is not found around the Earth. 
This difference comes from the difference in radial drift velocity.
A dust particle with larger $\beta$ radially drifts faster than that with smaller $\beta$.
The fraction of particles trapped in MMRs depends on $\beta$.
Almost all of particles with $\beta=0.001$ are trapped in MMRs with the Earth that produce a leading-trailing density asymmetry around the Earth, while very few particles with $\beta=0.2$ are trapped in MMRs.

The density profile around the Earth located at the perihelion is slightly different from that at the aphelion.
We use the surface number density of dust particles averaged over the Earth's orbit in following sections.

\subsection{Comparison of Observations and Calculations}
Using the number density $n$ derived in Section \ref{sec:dist}, 
we here calculate radially integrated radiative flux in the Earth-centric coordinate (see Section \ref{sec:obs}). 
We then obtain the ratio of the leading to trailing brightnesses using the results of orbital integrations.  

The ratio is almost independent of dust properties because the leading and trailing
brightnesses include almost same dust properties. Therefore, we simply
assume blackbody dust. The temperature profile is given by
\begin{eqnarray}
T=280\left( \frac{R}{1\,{\rm AU}} \right)^{-1/2} \ {\rm K}.
\end{eqnarray}
The intensity $I_{\nu}$ is calculated from
\begin{eqnarray}
I_{\nu}(r,\phi,\theta)=4\pi s^{2} \epsilon B_{\nu}(T) n(r,\phi,\theta),
\end{eqnarray}
where $B_{\nu}(T)$ is the Planck function and $\epsilon$ is the emissivity coefficient.
We assume $\epsilon=1$ because the size parameter $x \equiv 2\pi s/\lambda$ ($\lambda$ is an observation wavelength) is larger than unity.
In addition, since we focus on a ratio of the trailing to leading brightnesses, the effect of $\epsilon$ is almost cancelled out.

The radiative flux $F_{\nu}$ is calculated from the integration of $I_{\nu}$ as
\begin{eqnarray}
F_{\nu}=\int_0^\infty \int_{-\Delta \phi}^{\Delta \phi} \int_{\theta_{\rm l,t} -\Delta \theta_{\rm m}}^{\theta_{\rm l,t} +\Delta \theta_{\rm m}}
 \frac{I_{\nu}}{4\pi r^{2}} r^{2}drd\phi d\theta,
\end{eqnarray}
where $\theta_{\rm l,t}$ is the longitude in the leading or trailing direction, $\Delta \phi = 30^{\circ}$ is set for comparison with observations, and the model viewing angle $\Delta \theta_{\rm m} = 4.5^{\circ}$ is chosen as a moderately small value. 

\begin{figure}[h]
\begin{center}
\includegraphics[scale=0.5]{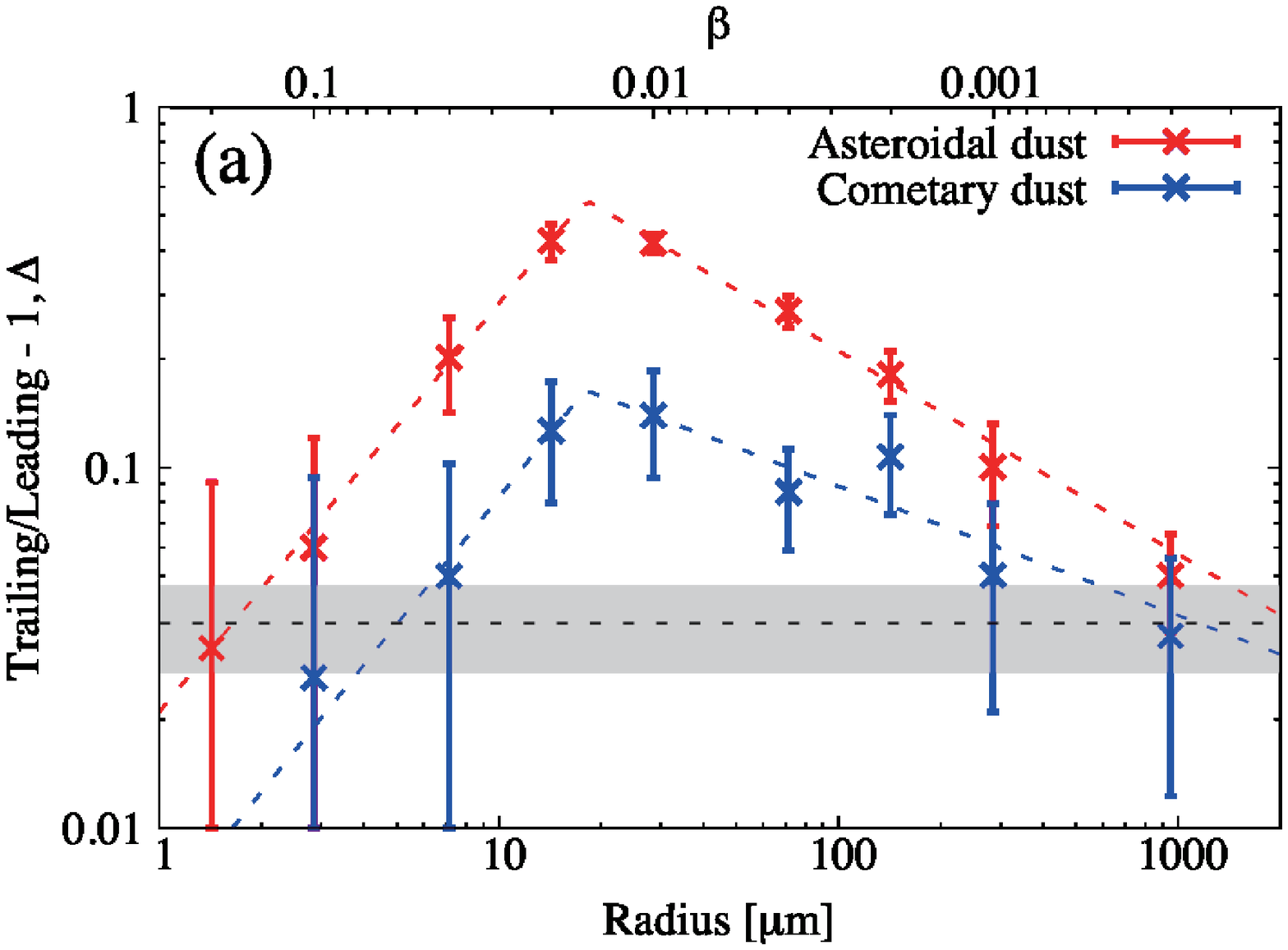}
\label{fig:18ratio}
\includegraphics[scale=0.5]{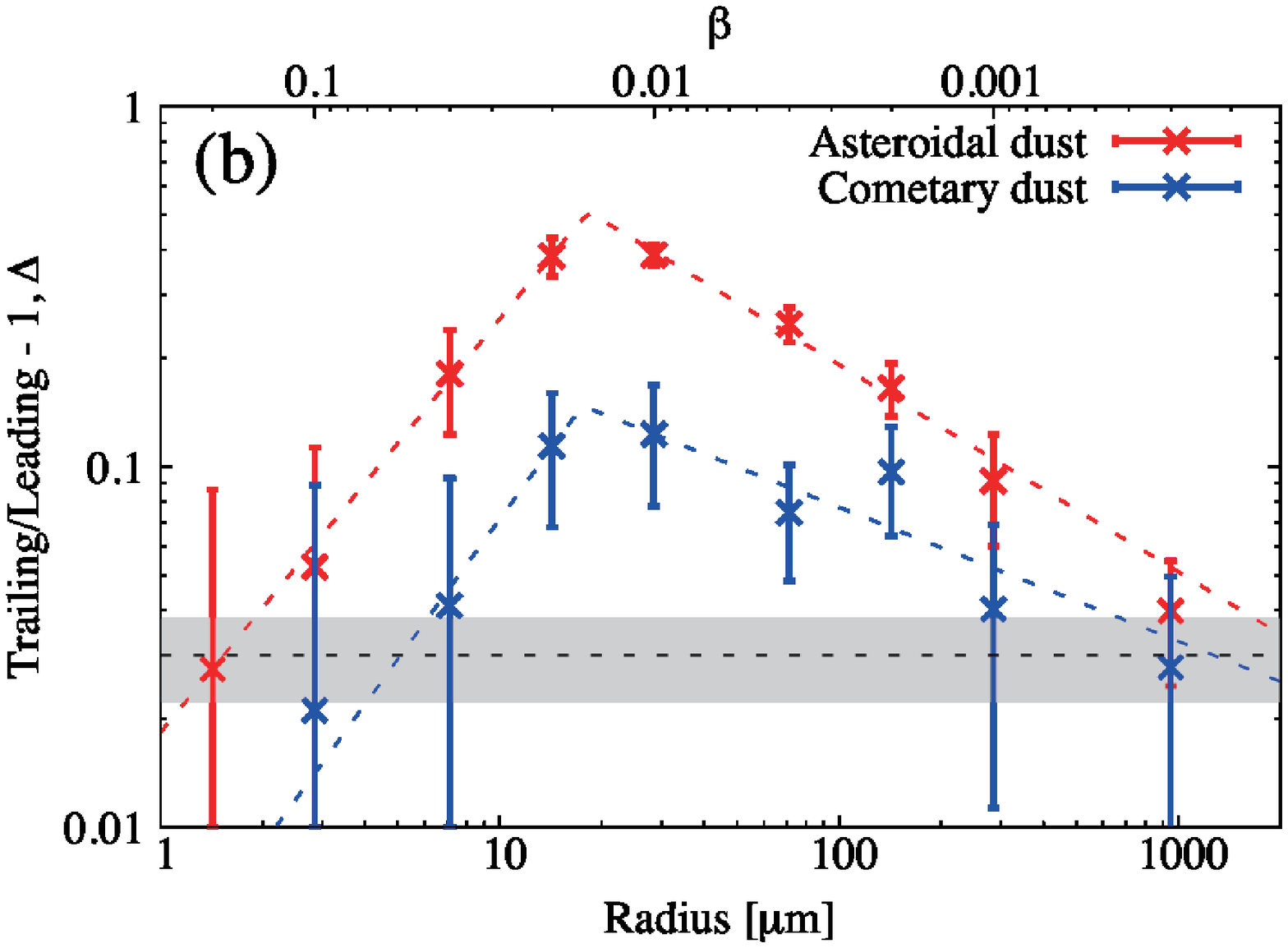}
\label{fig:9ratio}
\caption{
The surface brightness ratio of trailing to leading directions at wavelengths $9\,\micron$ (a) and $18\,\micron$ (b) as a function of the radius or $\beta$ of particles.
The relation between $s$ and $\beta$ is given with $\rho_{s}=2\,{\rm g/cm^{3}}$ and $Q_{\rm PR}=1$.
The red and blue points represent asteroidal and cometary cases, respectively.
The black dashed lines and gray filled regions correspond to the average value and standard deviation of observations, respectively, derived in Section \ref{sec:obs}.
The red and blue dashed lines denote the empirical formula for surface brightness difference, $\Delta$, used in Section \ref{sec:SFD}.
}
\label{fig:ratio-all}
\end{center}
\end{figure}

Figure \ref{fig:ratio-all} shows the surface brightness ratio of  the trailing to leading observations averaged over the orbital period of the Earth in the (a) $9\,\micron$ and (b) $18\,\micron$ bands for various dust sizes or $\beta$. 
The relation between $s$ and $\beta$ is given with $\rho_{s}=2\,{\rm g/cm^{3}}$ and $Q_{\rm PR}=1$.
We put error bars that are proportional to the inverse square root of the number of data points weighted by the square of the distance from the Earth.
The mean value obtained from the AKARI observations derived in Section \ref{sec:obs} is denoted by the black dashed line.
From Figure \ref{fig:ratio-all}, smaller particles with $s \lesssim 20\,\micron$ result in smaller difference in trailing and leading surface brightnesses.
This is caused by the probability of resonance trap that is low for smaller particles (i.e. larger $\beta$).
In contrast, for $s \gtrsim 20\,\micron$, the difference in trailing and leading surface brightnesses tends to decrease with increasing $s$.
This is mainly caused by two reasons.
The first one is the leading-trailing asymmetry caused by the orbital evolution of a single particle trapped in MMRs.
Larger particles (smaller $\beta$) trapped in MMRs produce  more symmetric distribution between the leading and trailing directions (see Figure \ref{fig:mmrs}).
The second one is that trapping MMR commensurability depends on particle size.
A particle with smaller $\beta$ is easier to be trapped in outer MMRs due to lower radial drift velocity.
The effect of outer MMRs on the leading-trailing difference in $F_{\nu}$ is smaller than that of inner MMRs because of larger distances from the Earth.

From Figure \ref{fig:ratio-all}, we conclude that asteroidal dust with $s \lesssim 3\,\micron$ and $s \ga 1000\,\micron$ and cometary dust with $s \lesssim 7\,\micron$ and $s \gtrsim 300\,\micron$ can explain AKARI observational asymmetries.
Meanwhile, the observational surface brightness ratio cannot be reproduced by the particle with $7\,\micron \lesssim \beta \lesssim 300\,\micron$ from asteroidal and cometary dust.

\section{Discussions}
\label{sec:discuss}

\subsection{The Effect of Size Frequency Distribution}
\label{sec:SFD}
Although single sized grains are considered in Section \ref{sec:orbital}, asteroidal and cometary grains have size frequency distributions. 
In order to investigate the size-integrated surface brightness differences, 
we obtain empirical formulas for surface brightness differences $\Delta_{s}\equiv F_{\rm trail}/F_{\rm lead}-1$ from the fitting of data shown in Figure \ref{fig:ratio-all}.
The empirical formulas are given by 
\begin{eqnarray}
\Delta_{s}=
\begin{cases}
A_{1}\left( \frac{s}{1\,\micron}\right)^{k_{1}} & {\rm for} \ \ s < \left( \frac{A_{1}}{A_{2}} \right)^{\frac{1}{k_{2}-k_{1}}}{\rm\mu m},\\
A_{2}\left( \frac{s}{1\,\micron}\right)^{k_{2}} & {\rm for} \ \ s > \left( \frac{A_{1}}{A_{2}} \right)^{\frac{1}{k_{2}-k_{1}}}{\rm\mu m}, \label{eq:emp}
\end{cases}
\end{eqnarray}
where $A_{1}, A_{2}, k_{1}$ and $k_{2}$ are constants and listed in Table \ref{tab:const}.
\begin{table*}
\begin{center}
\scalebox{1.0}{
\begin{tabular}{ccccc} \hline
constant & Asteroid (${\rm 9\,\mu m}$) & Comet (${\rm 9\,\mu m}$) & Asteroid (${\rm 18\,\mu m}$) & Comet (${\rm 18\,\mu m}$)\\
$A_{1}$& $0.02101$  & $0.005600$ & $0.01832$  & $0.003636$ \\
$k_{1}$ & $1.133$ & $1.169$ & $1.147$ & $1.294$ \\
$A_{2}$ & $2.823$ & $0.4609$ & $2.712$ & $0.4280$ \\ 
$k_{2}$ & $-0.5637$ & $-0.3579$ & $-0.5751$  & $-0.3720$ \\ \hline
\end{tabular}
}
\caption{The values for constants of Equations (\ref{eq:emp}).}
\label{tab:const}
\end{center}
\end{table*}
Using Equation (\ref{eq:emp}), the trailing and leading thermal fluxes emitted by particles with radius $s$, $F_{\rm trail}(s)$ and $F_{\rm lead}(s)$, respectively, are related as $F_{\rm trail}(s)=(1+\Delta_{s})F_{\rm lead}(s)$.
Because $F_{\rm lead}(s)$ is almost proportional to $s^{2}$, the size-integrated surface brightness differences $\Delta$ is given by
\begin{eqnarray}
\Delta=\frac{\int_{s_{\rm min}}^{s_{\rm max}}F_{\rm trail}(s) N_{\rm d}(s) ds}{\int_{s_{\rm min}}^{s_{\rm max}}F_{\rm lead}(s) N_{\rm d}(s) ds}-1
=\frac{\int_{s_{\rm min}}^{s_{\rm max}}s^{2}\Delta_{s}N_{\rm d}(s)ds}{\int_{s_{\rm min}}^{s_{\rm max}}s^{2}N_{\rm d}(s)ds}, \label{eq:delta}
\end{eqnarray}
where $N_{\rm d}(s)$ is the differential surface number density around the Earth orbit and $s_{\rm max}$ and $s_{\rm min}$ are the maximum and minimum dust radii, respectively.
We set $s_{\rm min}=0.57\,\micron$ corresponding to $\beta=0.5$ with $\rho_{s}=2\,{\rm g/cm^{3}}$ and $Q_{\rm PR}=1$.
We choose a sufficiently large value for $s_{\rm max}$, so that we set $s_{\rm max}=3000\,\micron$. 
In the main asteroid belt, the size distribution of dust particles is controlled by successive collisions or collisional cascade, 
resulting in the differential number density $n_{\rm d}(s)$ with a power-law spectrum.
If the collisional strength of particles is independent of $s$, $n_{\rm d}(s) \propto s^{-3.5}$ (\citealt{Dohnanyi1969}, \citealt{Tanaka1996}). 
Although the power-law index for realistic collisional strength is slightly different from $-3.5$ \citep{KT2010}, we simply consider $n_{\rm d}(s) \propto s^{-3.5}$.
The particles resulting from the collisional cascade have a modified size distribution due to radial drift. 
For 10 micron sized or smaller particles, collisional timescale $\tau_{\rm coll}$ ($\propto s^{0.5}$) is significantly longer than radial drift timescale $\tau_{\rm drift}$($\propto s$), so that dust particles produced by collisional fragmentation immediately drift inward.
In the steady state, the number of produced dust per unit time ($\propto n_{\rm d}(s)$) is equilibrated with the number flux of drifting dust. 
Therefore $N_{\rm d}(s)$ satisfies $2\pi Rv_{r}(s)N_{\rm d}(s) \propto n_{\rm d}(s)$, 
where $v_{r}(s)$ is the drift velocity and $R$ is the heliocentric distance. 
Since $v_r$ is roughly proportional to $1/s\,R$, the differential surface density around the Earth is given by $N_{\rm d}(s) \propto s^{-2.5}$.
For larger particles, since collisional timescale $\tau_{\rm coll}$($\propto s^{0.5}$) is shorter than radial drift timescale $\tau_{\rm
drift}$($\propto s$), only a fraction of dust grains, $\sim \tau_{\rm coll}/\tau_{\rm drift}$, can drift inward to the Sun. 
Therefore, $N_{\rm d}(s) \propto n_{\rm d}(s) \tau_{\rm coll}/\tau_{\rm drift}$, resulting in $N_{\rm d}(s) \propto s^{-4}$.
The slopes derived from these simple estimate roughly correspond to the model of \citet{Grun1985}, which can account for the dust fluxes measurements by spacecraft and the crater size distribution on the Moon.
The transitional size between regimes $N_{\rm d} \propto s^{-2.5}$ and $N_{\rm d} \propto s^{-4}$ is still uncertain.
According to \citet{Grun1985}, we set the transitional size to be $30\,\micron$.
From Equations (\ref{eq:emp}) and (\ref{eq:delta}) with this size distribution, the leading-trailing surface brightness asymmetry formed by asteroidal dust is obtained to be $\Delta=27.7\%$ at wavelength $9\,\micron$ and 25.3\% at $18\,\micron$.
The result is almost independent of the transitional size if the transitional size ranges from $10\,\micron$ to $100\,\micron$.
These values are approximately 8 times larger than the observational values; asteroid dust should have a minor contribution to the infrared zodiacal emission. .

Size distributions of cometary dust were measured by Stardust mission and Rosetta mission (e.g., \citealt{Horz2006}, \citealt{Moreno2016}).
The dust particles originating from comets have  $n_{\rm d} \propto s^{-\gamma}$ with constant $\gamma$.
Due to the size dependence of P-R drag discussed above, the differential size distribution of cometary dust around the Earth orbit satisfies $N_{\rm d}(s) \propto s^{-\gamma+1}$.
The dust particles originating from comet 81P/Wild 2 have $\gamma=2.72$ \citep{Horz2006}.
On the other hand, the dust particles on comet 67P/Churyumov-Gerasimenko have $\gamma=3$ \citep{Moreno2016}.
From Equations (\ref{eq:emp}) and (\ref{eq:delta}) with $N_{\rm d}(s) \propto s^{-\gamma+1}$, the leading-trailing surface brightness asymmetry formed by cometary dust is estimated to be $\Delta=3.6\%$ at wavelength $9\,\micron$ and  3.1\% at $18\,\micron$ if $s_{\rm max}=3000\,\micron$ for $\gamma=2.72$ and if $s_{\rm max}=4000\,\micron$ for $\gamma=3$.
These values are consistent with the values derived from AKARI observations.
Note that cometary values depend on the maximum dust radius $s_{\rm max}$ because the differential size distribution with $\gamma<4$ suggests that large particles dominate the total surface area.
These results suggest that maximum dust radius $s_{\rm max}$ had better to be larger than  $\sim 3000\,\micron$.
However, millimeter sized or larger grains on 67P/Churyumov-Gerasimenko have not $\gamma=3$ but $\gamma=4$ \citep{Moreno2016}.
If we assume $s_{\rm max}=1000\,\micron$ for $\gamma=3$, $\Delta=5.7\%$ at wavelength $9\,\micron$ and  4.9\% at $18\,\micron$, which are comparable to the maximum observational values.

Our calculation includes statistical errors with values of $\sim3\%$ (see Figure \ref{fig:ratio-all}). 
Therefore, it is difficult to exactly determine the proportion of asteroidal and cometary populations.
If we assume that the total surface area of zodiacal dust, which is roughly proportional to the mid-infrared brightness of zodiacal light, is composed of 10\% asteroidal particles and 90\% cometary particles, $\Delta=6.0\%$ at wavelength $9\,\micron$, which is allowed within the statistical errors of calculation.
It should be noted that if 90\% of the mid-infrared brightness mainly comes from cometary dust, the number density of small particles ($\lesssim 100\,\micron$) is dominated by asteroidal particles.
The dominant component of zodiacal dust may depends on its size; the measured flux by spacecraft may come from asteroidal dust, while the thermal emission is dominated by cometary dust of the order of $1000\,\micron$.

\subsection{The Effect of Perturbations from Other planets}
In Section \ref{sec:orbital}, the calculations do not include the perturbations from planets other than the Earth.  
However, the perturbations may increase orbital eccentricities of dust particles drifting to the Earth. 
Especially, cometary dust originating from Jupiter-Family comets might be strongly perturbed by Jupiter because of their orbital crossings with Jupiter.  
In order to investigate this effect, we conducted orbital integrations of cometary dust with 7 planets except for Mercury.  
We run orbital calculations for 1000 particles with $\beta=0.03$ that have the initial orbits same as Section \ref{sec:orbital}
(but the computational cost cut-off using Equation (\ref{eq:const}) is not applied to these calculations).  
As a result of calculations for $10^{6}$ years, 
695 particles migrating inward are completely orbitally decoupled with Jupiter (the aphelion distances of the particles are much smaller than perihelion distance of Jupiter), 
193 particles survive at outer region with high eccentricity (the average eccentricity is 0.74) and 
112 particles are scattered from their system ($e>1$).  
The average eccentricity of orbitally decoupled particles is 0.32 when the semi major axes are $3.5\,{\rm AU}$.  
This value is still similar to the average eccentricity of Jupiter-Family comets.  
This is because dust particles experience sufficiently numerous encounters with Jupiter and the eccentricity increase up to a value in dynamical equilibrium that should be comparable to that of Jupiter-Family comets. 
We also perform simulations for other values of $\beta$ and then find that the result is almost independent on the value of $\beta$.  
This is because the radial drift timescale of dust particles around Jupiter orbit are sufficiently longer than their synodic timescales with Jupiter.  
Therefore, the leading-trailing surface brightness difference calculated in Section \ref{sec:orbital} does not change even if we include the effect of perturbation from Jupiter.

\section{Summary}
\label{sec:summary}
First, we have analyzed the infrared all sky survey data from AKARI and
have found that observations in the trailing direction are 3.7(3)\% brighter than those in the leading direction in the $9(18)\,\micron$ band.  
These results are consistent with previous observations by IRAS \citep{Dermott1994}.  
Second, we have investigated orbital motion of asteroidal and cometary dust particles
considering the gravity of the Sun and the Earth and solar radiation.  
Orbital evolution of a dust particle is characterized by radiation parameter $\beta$ that represents the ratio of solar radiation pressure to solar gravity.  
We have found that dust particles with $\beta \sim 0.01$ (corresponding to $s \sim 30\,\micron$ with $\rho_{s}=2\,{\rm g/cm^{3}}$) produce a most significant leading-trailing surface brightness asymmetry due to resonance trap.  
Asteroidal dust with $s \lesssim 3\,\micron$ and $s \gtrsim 1000\,\micron$ and cometary dust with $s \lesssim 7\,\micron$ and $s \gtrsim 300\,\micron$ show leading-trailing brightness asymmetries comparable to that found by AKARI observations.  
On the other hand, leading-trailing asymmetries produced by particles with $7\,\micron \lesssim s \lesssim 300\,\micron$ coming from either asteroids or comets are too large to reproduce observational one.
The realistic size frequency distribution of asteroidal dust around the Earth orbit is considered as $N_{\rm d}(s) \propto s^{-2.5}$ for $s<30\,\micron$ and
$N_{\rm d}(s) \propto s^{-4}$ for $s>30\,\micron$, where $N_{\rm d}(s)ds$ is the surface number density of particles with radii from $s$ to $s+ds$. 
The leading-trailing brightness difference of asteroidal dust integrated over the size distribution is obtained to be 27.7\% at wavelength $9 \mu {\rm m}$ and 25.3\% at $18\,\micron$. 
On the other hand, the size frequency distribution of cometary dust around the Earth orbit based on the measurement by the stardust mission is given by $N_{\rm d}(s) \propto s^{-\gamma+1}$ with $\gamma=2.72$ \citep{Horz2006}.
For cometary dust with the size distribution, the leading-trailing difference is 3.6\% at wavelength $9\,\micron$ and 3.1\% at $18\,\micron$ with maximum dust radius $s_{\rm max}=3000\,\micron$.
From these values and the errors, we have concluded that the population of $100$--$1000\,\micron$ sized zodiacal dust
includes less than 10\% asteroidal dust and more than 90\% cometary dust, and the maximum radius of cometary dust is as large as $1000\,\micron$.

\acknowledgments
The authors  acknowledge Shigeru Ida and Takafumi Ootsubo for fruitful discussions and comments.
This work was supported by Grants-in-Aid for Scientific Research (No. 26287101) from Ministry of Education, Culture, Sports, Science and Technology (MEXT) of Japan and by Astrobiology Center Project of the National Institute of Natural Science (NINS) (Grant Number AB281018).


\bibliographystyle{./aasjournal}
\bibliography{./zod}

\end{document}